\documentclass{pas}

\usepackage{multirow,aas-macros}
\usepackage{amssymb,natbib}
\usepackage{amsmath}

\begin{document}

\lefttitle{Publications of the Astronomical Society of Australia}
\righttitle{Cambridge Author}

\jnlPage{1}{4}
\jnlDoiYr{2021}
\doival{10.1017/pasa.xxxx.xx}

\articletitt{Research Paper}

\title{Flip-flop states in X-ray binaries and changing-state AGN}


\author{\sn{Thomas J.} \gn{Maccarone}$^{1}$, \sn{Jessie} \gn{Runnoe}$^{2}$, \sn{Gregoire} \gn{Marcel}$^3$, \sn{Emilia} \gn{J\"arvel\"a}$^{1}$,\sn{Douglas} \gn{Buisson}$^4$, \sn{Unnati} \gn{Kashyap}$^1$, \sn{Federico M.} \gn{Vincentelli}$^{5,6,7}$  }
\affil{$^1$Department of Physics \& Astronomy, Texas Tech University, Lubbock TX 79409-1051, $^2$ {Department of Physics \& Astronomy, Vanderbilt University,2401 Vanderbilt Place, Nashville, TN, 37240, USA}, $^3$ {Department of Physics and Astronomy, FI-20014 University of Turku, Finland}, $^4$ {Independent scientist}, $^5${Fluid and Complex Systems Centre, Coventry University, CV1 5FB, UK}, $^6$ {INAF—Istituto di Astrofisica e Planetologia Spaziali, Via del Fosso del Cavaliere 100, I-00133 Roma, Italy}, $^7${School of Physics \& Astronomy, University of Southampton, Southampton SO17 1BJ, UK}}

%


\corresp{T.J. Maccarone, Email: thomas.maccarone@ttu.edu}


\history{(Received xx xx xxxx; revised xx xx xxxx; accepted xx xx xxxx)}

\begin{abstract}
We show that the flip-flop transitions in X-ray binaries {(rapid cycling between different spectral states which are sometimes seen near the global state transition)} show a series of analogies to the changing state phenomena { (rapid changes in the emission line properties that seem to be driven by changes in the central engine)} in active galactic nuclei (AGN).  Specifically, (1) the timescales for the transitions scale approximately linearly with mass and (2) both phenomena occur at a few percent of the Eddington luminosity.  { Because most accretion physics is expected to be scale-free, it is likely that these represent two manifestations of the same phenomena. Demonstrating this would allow the use of a much wider range of observational techniques, on a much wider range of characteristic timescales, and provide a clearer pathway toward understanding these rapid transitions than is currently available.}  We discuss potential means to establish the connection more firmly, and {to} use the combination of the observational advantages of both classes of systems to develop a better understanding of the phenomenon.
\end{abstract}


\maketitle

\section{Introduction}

Black hole accretion is a fundamentally important astrophysical process.  On the supermassive scale, it powers active galactic nuclei, which are one of the main sources of energetic feedback that affects cosmological structure formation  \citep{FabianFeedback}.  On the stellar mass side, it also provides the means for identifying stellar mass black holes in environments where their spins can be estimated, giving clues to the nature of the supernova explosion mechanism that cannot be achieved any other way  \citep{Fryer}.  In both cases, these systems may also be outstanding particle accelerators, producing high-energy photons  \citep{Punch, Tavani}, neutrinos \citep{IceCube,Koljonen} and cosmic rays  \citep{HeinzSunyaev,Auger}.

{ Most phenomena related to black hole accretion are expected to be relatively scale-free, in the sense that accretion geometry and kinetic power into the jet will depend primarily on the accretion rate scaled to the black hole mass; only a few aspects of black hole accretion have strong dependencies on the density and temperature of the accretion flow because it affects the atomic physics equilibria.  Indeed, past work has shown many analogies in the behavior of black hole X-ray binaries and AGN.  For example, it was first seen in X-ray binaries that there are some sharp transitions in their X-ray spectral and timing behaviors that can set on relatively quickly, indicative of a fundamental change of the geometry of the accretion flow \citep{Tananbaum1972}. Evidence for spectral states similar to those of stellar mass black holes appears in power spectra  \citep{McHardy2006}, jet power  \citep{Tananbaum1972,Merloni2003,MaccaroneGalloFender}, and spectral energy distributions  \citep{Tananbaum1972,Ho1999,Trichas2013}, and the transition between phenomenology similar to that in the hard state and similar to that in the soft state occurs at luminosities close to 2\% of the Eddington limit in both X-ray binaries  \citep{MaccaroneStates,2019MNRAS.485.2744V} and AGN  \citep{MaccaroneGalloFender,2006MNRAS.372.1366K,Moravec}. 

For studies of accretion physics, stellar mass and supermassive black holes provide highly complementary data.  For most purposes, stellar mass black holes, with their much higher signal-to-noise, can be studied in greater detail.  Their black hole masses are more straightforward to measure, at least in the cases of transients with bright donor stars.  Because of their smaller masses and faster characteristic timescales, black holes in transient accretion discs can vary by factors of millions within years, allowing a system with a constant mass and spin to be studied over a wide range of accretion rates.  Additionally, because of their higher disc temperatures, they are closer to fully ionised, making the physics of their accretion discs simpler to understand.  Such studies of stellar mass black holes can then inform the best way to assemble samples of AGN to probe their accretion physics.

Supermassive black holes have a few key advantages, as well.  In particular, for certain purposes, their longer characteristic timescales are a benefit.  In supermassive black holes, phenomena happening on thermal, and even dynamical, timescales can be seen directly evolving.  In stellar mass black holes, these phenomena can often be studied only statistically, via Fourier methods.   In particular, stellar mass black holes typically show count rates of a few thousand counts per second, with dynamical timescales of a few milliseconds, yielding of order 1-10 photons per dynamical timescale with high collecting area facilities like the Rossi X-ray Timing Explorer and Neutron star Interior Composition Explorer (NICER).  The nearest bright AGN show count rates a few orders of magnitude smaller, but with characteristic timescales about 5-7 orders of magnitude longer, so that there are often thousands of counts per dynamical timescale.  The fastest variability, in units of the dynamical timescale, is thus best probed in AGN (e.g. \citealt{McHardy2006}).  Fourier methods do allow compensating for the lack of photons per characteristic timescales in the stellar mass black holes by making statistical studies possible over large numbers of characteristic timescales. Table \ref{table_timescales} shows some characteristic timescales for accretion discs around different black hole masses.

Additionally, supermassive black hole accretion discs emit more of their power in the optical band. Because optical telescopes are far more sensitive in flux units than X-ray telescopes, they allow the discovery and time-domain monitoring of a much larger number of AGN than of stellar mass black holes (e.g. \citealt{2018ApJ...854..160R,2024SerAJ.209....1K}).  This can even include spectroscopic monitoring \citep{2025ApJ...986..160D}.  As a result, intrinsically rapid variability can be discovered in AGN from survey data, while in stellar mass black holes, it can generally only be discovered in pointed observations.
}

One of the key findings from studies of stellar mass black holes is that they show a set of states in which spectral properties, variability properties, and relativistic jet power are strongly correlated.  The ``hard state" is characterised by a $\Gamma=1.5-1.8$ power law with a rollover around 100 keV (along with a weak thermal accretion disc peaking below 1 keV), broadband variability with a { root mean square (rms)} amplitude of up to about 30\%  \citep{vanderklis1994}, and with strong radio emission that usually follows a $L_R\propto L_X^{0.7}$ relation  \citep{1998A&A...337..460H,2003MNRAS.344...60G}.  The soft states have properties that agree well with the predictions of the Shakura-Sunyaev disc model, and are well-matched spectrally as sums of blackbodies with $T\propto R^{-3/4}$  \citep{Zhang1997}, and show little to no variability, and generally show no radio emission  \citep{Tananbaum1972,Maccarone2020}.  Intermediate states show a rather complicated phenomenology, with spectra that generally have strong thermal and strong non-thermal components, and power spectra that show strong quasi-periodic oscillations \citep[QPOs, see, e.g.,][]{Motch1983,2001ApJS..132..377H}.

In recent years, both the X-ray binaries and the AGN have shown some examples of objects that change spectral appearances very quickly.  In the X-ray binaries, these changes are known as flip-flop transitions  \citep{Takizawa1997}.  In the AGN, the objects are called changing-look AGN  \citep{RicciTrakhtenbrot}.  In this paper, we discuss the broad similarities of the two phenomena, and discuss potential observational tests that could be used to determine if they are, indeed, two manifestations of the same process, on different mass scales.

\section{Properties of flip-flop transitions}
The first clear example of flip-flop transitions in X-ray binaries was GX~339-4  \citep{Miyamoto1991}.  In these observations, the source count rate alternates between two distinct levels, differing by only a few percent. These transitions occur on timescales of seconds, yet a given state can persist for hours. In conjunction with these changes, flip-flops see the variability shift between two markedly different regimes (typically between cases dominated by a single QPO and cases dominated by broadband aperiodic variability). For a comprehensive summary of the literature on this topic, we refer the reader to the introduction of  \cite{Buisson2025}.  

The flip-flop transitions can be seen both in hard-to-soft state transitions, and in soft-to-hard transitions  \citep{Buisson2025}.  These generally occur at somewhat different Eddington fractions, because of hysteresis in state transition luminosities  \citep{Miyamoto1991,2003MNRAS.338..189M}.  It is likely that if changing-state AGN { (CSAGN)} are the analogs of flip-flop transitions, and AGN follow the same outburst phenomenology with hysteresis effects  \citep{2006MNRAS.372.1366K} that the CSAGN will be seen mostly during the slower, fainter soft-to-hard state transitions, simply because those last longer.

Normally, state transitions are thought to be driven by changes in accretion rate.  The viscous timescale, $t_{visc}$, in an accretion disc is thought to be the fastest timescale on which the accretion rate can vary substantially.
It is given by:
\begin{equation}
    t_{visc}= \alpha^{-1}\left(\frac{H}{R}\right)^{-2} t_{dyn},
\end{equation}
where $\alpha$ is the dimensionless viscosity parameter, $H$ is the vertical height of the disc, $R$ is the radial scale of the disc for which the timescale is calculated, and $t_{dyn}$ is the dynamical timescale, which is the Keplerian orbital period.

For typical parameters for X-ray binaries in states dominated by geometrically thin components, this timescale will be typically a few seconds or more (although if a very small region mediates the state changes, the viscous timescale across that region may be small --  \citealt{2020A&A...641A.167S}).  The flip-flop transitions thus set on with timescales much faster than the local viscous timescale, indicating that they are {\it not} likely due to fundamental changes in the mass transfer rate; in particular because one typically expects variations in accretion rate to be driven by the outer regions, where $t_{visc}$ is of the order of hours.  Also in agreement with this is that there are no large changes in the bolometric luminosity across the flip-flop transitions.

An alternative timescale of interest is the thermal timescale, $t_{th}$ (see e.g.  \citealt{Bogensberger}).  This is given by:
\begin{equation}
    t_{th} = \alpha^{-1} t_{dyn}
\end{equation}
Contrarily to the (local) viscous timescale, this thermal timescale can be around or below a second, i.e. the right order of magnitude for spectral changes observed during flip-flops of X-ray binaries  \citep{Buisson2025}.

\begin{table}[h!]
\centering
\begin{tabular}{lccc}
\hline\hline
\textbf{disc Timescales} 
& $10\,M_\odot$ 
& $10^7\,M_\odot$ 
& $10^{10}\,M_\odot$ \\
\hline
$t_{\rm dyn} = \Omega_K^{-1}$ 
& $1.6\,\mathrm{ms}$ 
& $26\,\mathrm{min}$ 
& $18\,\mathrm{days}$ \\[4pt]

$t_{\rm th} = \alpha^{-1} t_{\rm dyn}$ 
& $16\,\mathrm{ms}$ 
& $4.3\,\mathrm{h}$ 
& $181\,\mathrm{days}$ \\[4pt]

$t_{\rm visc} = \alpha^{-1}(H/R)^{-2} t_{\rm dyn}$ 
& $1.6\,\mathrm{s}$ 
& $18\,\mathrm{days}$ 
& $49\,\mathrm{yr}$ \\
\hline\hline
\textbf{State timescales} 
& $10\,M_\odot$ 
& $10^7\,M_\odot$ 
& $10^{10}\,M_\odot$ \\
\hline
X-ray state transitions 
& $\gtrsim 1\,\mathrm{week}$ 
& $\gtrsim 20\,\mathrm{kyr}$ 
& $\gtrsim 200\,\mathrm{Myr}$ \\[4pt]


Flip--flop state transition
& s to hours 
& wks to centuries
& $30\,\mathrm{years}$ to Myr \\
\hline
\end{tabular}
\caption{(Top) Theoretical disc timescales at $R=10\,R_g$ assuming  $\alpha=0.1$ and $H/R=0.1$. (Bottom) Observed characteristic timescales of stellar-mass black hole transitions and their mass-scaled AGN equivalents (assuming $t \propto M$).}
\label{table_timescales}
\end{table}

\begin{figure*}
    \centering
    \includegraphics[width=0.8\linewidth]{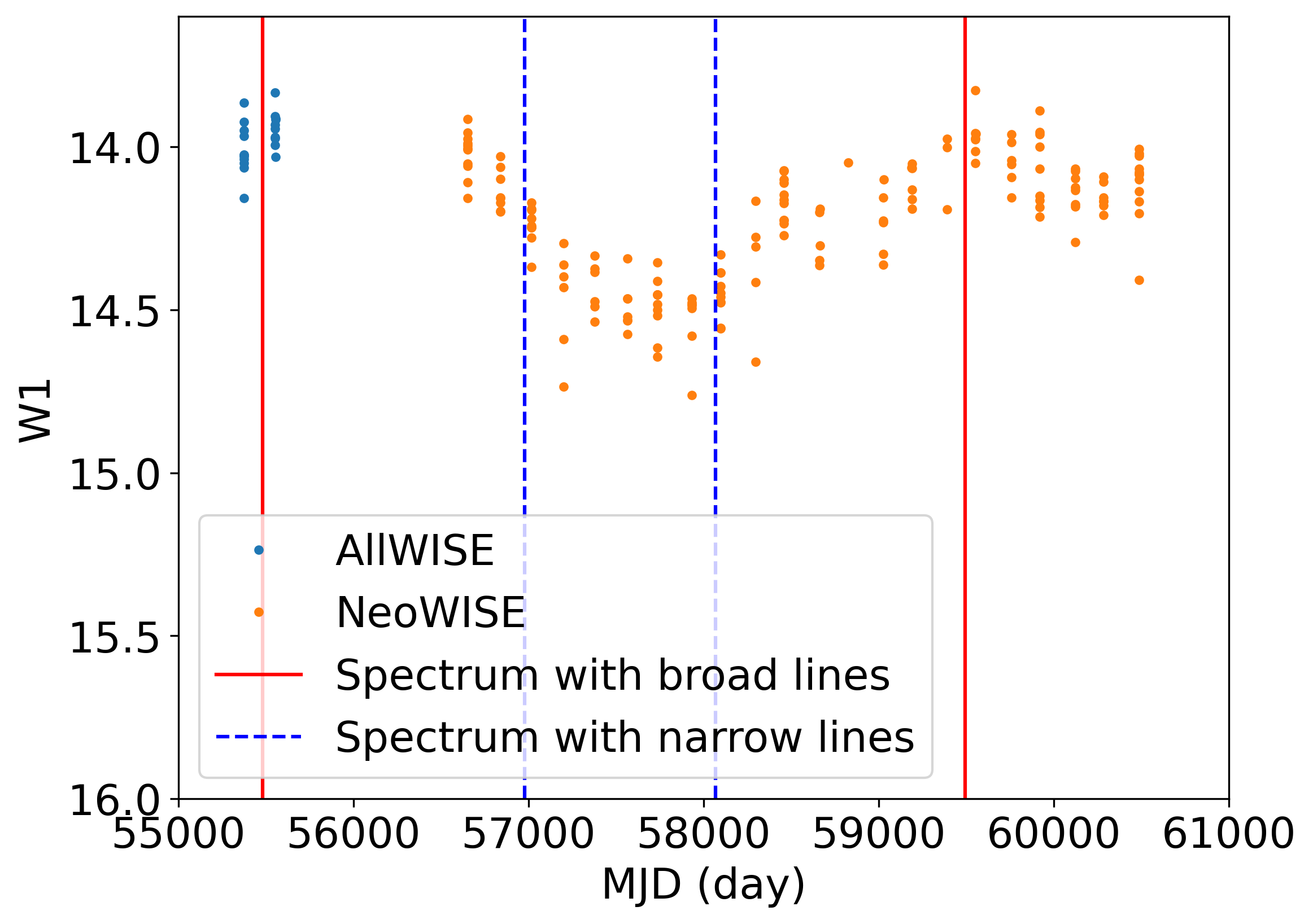}
    \includegraphics[width=0.8\linewidth]{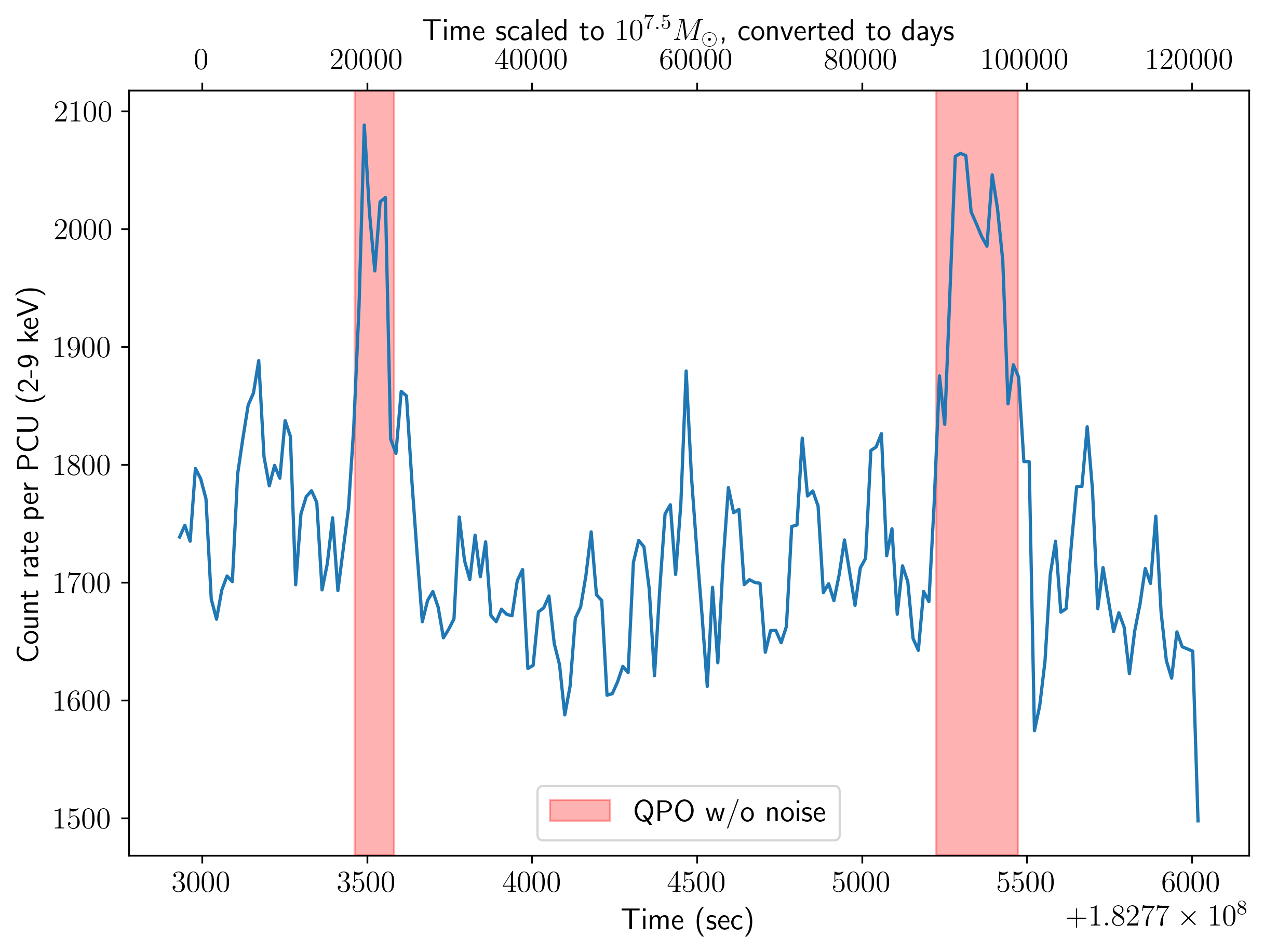}
    \caption{Top: the WISE light curve for the recurring CSAGN, J002311.06+003517.5, reported in \cite{wang25,DongLAMOSTsample}, chosen because it shows the relevant state change and return most clearly of the recurring state change objects.  For the black hole mass of approximately $10^9 M_\odot$, the faint state lasts the equivalent of about 3 seconds for a stellar mass black hole, and is one of the fastest objects in the sample of \cite{DongLAMOSTsample} in terms of during relative to black hole mass.  Bottom: RXTE data from the source XTE~J1859+226, first reported in \cite{2004A&A...426..587C}.  Here, sharp changes in count rate are seen in association with changes in the Fourier power spectrum; in the fainter states, there is strong aperiodic variability and no clear QPO, while in the brighter states, there is a clear QPO and little or no aperiodic variability.  The upper axis label gives the time in units re-scaled to a typical CSAGN with a black hole mass of $10^{7.5} M_\odot$. }
    \label{fig:analogyfigure}
\end{figure*}

\section{Properties of changing look AGN}

 \citet{RicciTrakhtenbrot} give a recent review of the details of changing-look AGN (CLAGN).  They appear to be a heterogeneous class, with a subset of CLAGN caused by changing absorption, and the remainder being due to some fundamental change in the accretion process.  These are sometimes classified in a manner that distinguishes between changing obscuration and changing state.

In a substantial sample of objects, there is strong evidence for a pattern in which the CSAGN behave in a manner similar to changes of states in X-ray binaries.  Specifically, the UV to X-ray spectral slopes change at Eddington fractions similar to those for the X-ray binaries \citep{2018MNRAS.480.3898N,Ruan2019,2024ApJS..272...13P}.  Samples of steadier AGN also show phenomenology indicative of spectral states changing at such Eddington fractions \citep{MaccaroneGalloFender,2011MNRAS.413.2259S}.  In several cases, there is even clearer evidence that a change in accretion disc intrinsic properties, rather than a change in absorption must be responsible for the changes in CSAGN  \citep{Duffy2025}.  The observed spectral changes are more dramatic than for the X-ray binaries, but this is likely because in AGN the thermal disc emision is in the optical-through-UV range, well separated from the Compton scattering emission.

The characteristic timescales for the CSAGN (Table \ref{table_timescales}) appear to be roughly consistent with what would be expected from scaling up the stellar mass black holes' flip-flop oscillations, based on a few objects seen to return from faint states to bright ones  \citep{Duffy2025b, DongLAMOSTsample}.  The source SDSS~J101152.98+544206.4 took about 500 days to make its state transitions, and stayed in a faint state for about 10.6 years.  Given its black hole mass of $10^{7.6} M_\odot$, assuming a linear scaling of key timescales with mass, this corresponds to an ingress time of a few seconds for a stellar mass black hole and a faint state duration of about a minute  \citep{Duffy2025b}; Mrk 1018, which has a black hole mass a factor of about three higher had a bright state a factor of about three longer  \citep{Dunn2025}.  The two other sources in  \cite{Duffy2025b}, both of which showed the same phenomenology of returning to bright states, were less well sampled and have larger mass black holes, but are broadly consistent with the same timescales.  In a larger sample of objects with LAMOST spectroscopy, about 10\% of objects showed recurring changes, with typical observed recurrence times of a few years  \citep{DongLAMOSTsample}.  Here, again with relatively sparse sampling meaning that it is likely that many objects had missed transitions, so that some recurrence times may be faster than a few years, and a higher percentage of objects likely show recurring changes within timescales of a few years.  In comparison,  \cite{Buisson2025} finds state changes with typical durations of tens of seconds in X-ray binaries (see e.g. their Figure 6).  Better sampling of more AGN is needed to develop conclusive evidence that these timescales are proportional to black hole mass, but the existing evidence is strongly suggestive of this.

\section{Potential tests and applications of the scenario}

Broad similarities exist between the flip-flop state transitions in X-ray binaries and the CSAGN transitions.  Specifically, (1) the characteristic timescales of the ingresses, egresses, and durations of the states generally are consistent with being linearly proportional to the black hole masses; { this is illustrated in figure \ref{fig:analogyfigure},} (2) the spectral shape changes are consistent with changes between a roughly Shakura-Sunyaev type accretion disc and { an advection dominated accretion flow-}like hard spectral state and (3) the phenomenon occurs at luminosities near the few percent of the Eddington limit  \citep{Ruan2019,2024ApJS..272...13P} where stellar mass black holes show their spectral state transitions. Having established these broad similarities between the phenomena of flip-flop transitions in black hole X-ray binaries and CLAGN, we now consider observational tests of the picture.

\subsection{Searches for QPOs in CSAGN}
In the X-ray binaries, the flip-flop states are often associated with the presence of low frequency (i.e. few Hz) QPOs  \citep{Buisson2025}.   Assuming the periods of the oscillations scale linearly with black hole mass, they should be at periods of $P\approx$2 days~$\left(\frac{M_{BH}}{10^7 M_\odot}\right)$ in AGN.  

The similarities between the spectral states indicate that these would be good places to look for QPOs in AGN.  In stellar mass black holes, the amplitudes of the QPOs are seen as much higher in harder X-ray bands, dominated by the power law components  \citep{BelloniQPOhardness}.  For the nearest systems, observations with the Swift { X-ray Telescope} of about 5 ksec each are sensitive enough to detect these systems.  The exposure times needed are thus quite substantial, as about 100 observations would be needed to detect the few tens of QPO cycles needed to distinguish from red noise, but this is achievable.  With future instruments { like NICER} with better angular resolution and hence better characterised background systematics (e.g. as proposed for STROBE-X) this could be done in a significantly shorter exposure time  \citep{Ray2024}.  It could also be done with an imager with more collecting area than Swift, but similar responsiveness, as proposed for AXIS  \citep{AXISPaper}.  

We note that one QPO has been seen in X-ray data from a CSAGN  \citep{2025Natur.638..370M}.  The phenomenology of this QPO does not match well to that of QPOs seen from stellar mass black holes in the past  \citep{2025Natur.638..370M}.  In particular, it shows characteristic frequencies that would scale to a few hundred Hz if the frequency scales linearly with the black hole mass, and shows clear evolution to higher frequencies with time.  

{ Whether QPOs are present, but undetected in other CSAGN is unclear.  The AGN in which \cite{2025Natur.638..370M} found the QPO is one of the closest CSAGN, and attracted significant attention and telescope time due to having had a major outburst.  Furthermore, it has a relatively low mass black hole for an AGN.  The QPO also fits within a single XMM observation's duration, rather than requiring monitoring to be detected.}

The most comprehensive search for high frequency QPOs in stellar mass black holes  \citep{2012MNRAS.426.1701B} was done (1) with power spectra averaged over full RXTE observations, so that oscillations with strong frequency evolution would change and (2) by searching only the range from 100-1000 Hz, so that if there were a very broad excess of Fourier power around a few hundred Hz due to an oscillation changing frequency with time, it would not present a strong contrast relative to adjacent frequencies.  As a result, searches for QPOs analogous to the ones seen in  \citet{2025Natur.638..370M} may be warranted during the flip-flop transition states.

{ It is important to note that the QPO phenomenology is rather complicated for the flip-flop transitions, and not yet fully understood.  Whether the QPOs are seen in the bright or the faint states varies from source to source \citep{Bogensberger}, and in some cases, multiple different variability patterns are seen in conjunction with the QPOs \citep{2004A&A...426..587C}. At the present time, it is probably most fruitful simply to search for QPOs in the light curves of the CSAGN, and if they are found, to use them in conjunction with the properties of the QPOs from flip-flopping X-ray binaries to try to develop a real understanding of the nature of the transitions that are happening.}

\subsubsection{Prospects for future detections}

{ For what we expect to see more typically (rather than what was seen in \citealt{2025Natur.638..370M}), QPOs on timescales of a few days, one needs sub-Nyquist sampling of the sources (i.e. observations typically about twice a day).  One also needs to sample enough cycles of the oscillation to be sure that one is not seeing red noise masquerading as a QPO, a known serious problem for AGN variability \citep{2016MNRAS.461.3145V}.  To make an estimate of the exposure time needed to detect such a phenomenon, we can look at the signal-to-noise for seeing quasi-periodic oscillations from \citet{2000ARA&A..38..717V}: $n_\sigma=\frac{1}{2}Ir^2(T/\lambda)^{1/2}$, where $n_\sigma$ is the signal to noise, $I$ is the count rate, $r$ is the rms amplitude of the oscillation, $T$ is the exposure time and $\lambda$ is the frequency width of the oscillation.  We can then re-arrange this for values typical of the expectations for a supermassive black hole's QPO and solve for the needed exposure time: $T = 10^6 {\rm sec} (n_\sigma/5)^2 (\lambda/10^{-6} {\rm Hz}) (r/0.03)^{-4} (I/10^{-2} {\rm cts/sec})^{-2} $.  The equation assumes sufficiently good sampling to resolve the QPO in time, and a sufficiently long light curve to cover the $\approx$ 20 cycles of the oscillation needed to ensure that the phenomenon is not red noise.  Such light curves may be achievable at the present time with Swift XRT monitoring for some of the brighter objects with lower black hole masses, like NGC~4151.}

It is also possible that high cadence, high sensitivity optical monitoring could detect the QPOs, but this would likely be from thermal reprocessing, rather than direct emission, and would be at lower { rms} amplitude  \citep{edelson2019,cackett2021}.  Still, given the relative ease of obtaining high cadence optical data relative to high cadence X-ray data, this may be a practical approach to searching for QPOs.  In the case of AGN, this may be possible in the optical  \citep{2018ApJ...860L..10S}. { This could be done best with facilities like the Argus Array  \citep{Argus} taking advantage of its excellent cadence. }

{ The amplitudes of modulation that might be expected in the optical band are probably rather small, as the QPOs in the X-ray binaries show much stronger amplitudes in energy ranges dominated by non-thermal coronal emission than thermal disc emission, with strengths of the particular class of QPO (Type B) that is most often seen during flip-flop transitions of about 0.5\% rms amplitude in the disc-dominated energy bands.  For comparing with optical data sets, we take a re-arrangement of the QPO signal-to-noise formula where $n_\sigma = \frac{1}{2} n_{\sigma,det}^2 r'^2 (\lambda T)^{-0.5}$ \citep{2019RNAAS...3..116M}, where $n_{\sigma,det}$ is the signal-to-noise for the detection of a source in the full set of observations used to make the light curve, and $r'$ is the intrinsic variability amplitude of the source (which, in background-limited data, may be much larger than the amplitude observed).  $T$ remains the actual exposure time for intermittently sampled light curves, not the duration of the time series.  

We can then consider a case where a light curve of 3 months duration is produced, with about 10\% duty cycle on source (yielding $T=7.5\times10^5$ seconds), to search for a 2-day period QPO (i.e. $5.8\times 10^{-6}$ Hz) with frequency width of $10^{-6}$ Hz.  We find that we need $n_{\sigma,det} = 600 n_\sigma^{1/2} (r'/0.005)^{-1}(\lambda T/0.75)^{1/4}$ to make a QPO detection.  The Argus Array should reach a magnitude of about 26 for $5\sigma$ detections \footnote{https://argus.unc.edu/specifications} in 3 months, while $n_{\sigma,det}$ of 600 should be reached for objects about 5 magnitudes brighter.  We can then have some reasonable hope of detection optical QPOs with Argus for AGN brighter than about 21st magnitude as long as the black hole masses are in the vicinity of $10^7 m_\odot$ or smaller, but we also emphasize that this prediction is based on extrapolations that are rather uncertain.  Roughly half the sky should have Argus coverage for at least three months per year.  We can thus expect that essentially all of the CSAGN seen from the Sloan Digital Sky Survey \citep{RicciTrakhtenbrot} would be bright enough to see QPOs if our baseline assumptions are correct.  Some additional work may be possible with the Vera Rubin Observatory's Legacy Survey of Space and Time \citep{LSST}, which will be able to detect objects a factor of a few fainter at the expense of cadence typically of 3 days, limiting us to searches for QPOs from AGN of more than about $10^8 M_\odot$.  

Finally, as a caution, we emphasize that it} is not clear that QPOs should be detectable in all of these systems, even if the same phenomenology is responsible for flip-flop transitions and CSAGN.  First, there appears to be a modest inclination angle dependence for the amplitudes of different classes of  QPOs  \citep{Motta2015}.  Second, some of the models invoked to explain the properties of QPOs require a misalignment between the black hole spin axis and the orbital plane of the binary, in order to allow for frame-dragging leading to a Lense-Thirring precession \citep{StellaVietri,Motta2018}. Some early work suggested that large misalignments of radio jet axes from the axes of inner dust discs thought to trace the accretion disc axis were small  \citep{1979A&A....73L...1K}, but more recent work indicates that misalignments are likely  \citep{Schmitt2002,Ruffa2020}. Alternative models can explain inclination angle dependent amplitudes without invoking misaligned discs  \citep{Varniere2017}, and furthermore, the misalignment angles can be modest ($\sim 15$ degrees) while still producing the QPOs  \citep{Ingram2009}.  Regardless of the potential complications, the CSAGN present excellent opportunities to search for QPOs, as they are as likely to show them as any class of AGN, and if QPOs are detected, they will be very illuminating about the nature of the changing state phenomenon.

\subsection{Searches for rapid cycling of jet emission}

Another core potential observational test relates to the properties of the relativistic jets.  In the X-ray binaries, steady jet production is seen in hard states, but not in soft states  \citep{Tananbaum1972}.  The steady jets are thought to convert their bulk kinetic energy into radiation through internal shocks  \citep{Spada,Jamil}.  

Cross-correlations between X-ray (which comes from the accretion inflow) and infrared (which comes from the jet) light curves in GX~339-4 show a correlation coefficient of about 0.3 at their peaks, with the infrared lagging behind the X-rays by about 0.1 seconds  \citep{2010MNRAS.404L..21C} -- the correlation is clearly statistically significant, but is also clearly not perfect.  This can be explained by the internal shock model if there is a correlation between the jet speed and the X-ray luminosity  \citep{2018MNRAS.480.2054M}. 

The essence of the issue here is that in the context of the internal shock model  \citep{2018MNRAS.480.2054M}, the observed {\it emission} from the jet on short timescales is determined by the {\it history} of the mass and speed of material injected into the jet.  In standard hard states where this model has been tested, the X-ray variability is up to about 30\% rms amplitude, with most of the variability power on timescales of seconds and faster.  In flip-flop states, we can expect a rather different scenario than a steady hard state, with a lot of mass and power injected into the jet during the hard states, and little or no power injected in the soft states.  

For wavelengths of light which lag the X-rays by very short timescales (such as the infrared), we can expect very little change in the observed variability properties.  Starting at time $\tau$ after the state transition sets in, the properties of the previous state's jet cease to be relevant.   At much longer wavelengths (such as the centimeter band), where $\tau$ exceeds the duration of the flip-flop's state change by a large factor, the effects of the flip-flopping will be smeared out to the point that they are not easily recognised.  In these bands, we can expect to see somewhat weaker radio emission, because the duty cycle of the jet is reduced, and we can expect to see radio emission during both the soft and hard states, but we do not expect to see dramatic changes to the variability properties.

At intermediate wavelength, $\lambda_{int}$, for which $\tau$ has a value of a few minutes, similar to the duration of a soft or hard state during a flip-flop transition, we may expect rather dramatic changes to the jet-band light curves.  In these cases, when a fast shell of material reaches the zone where photons at $\lambda_{int}$ are usually emitted, in some cases there will be no matter with which to interact, because the jet has been off for a while.  On the other hand, when that shell of material reaches the ``old jet" matter, it will not yet have dissipated much of its energy, so it will be moving faster, and have more energy to finally release.  This will lead to stronger variability in these wavelengths.  From  \cite{2010MNRAS.404L..21C} we can see that, for X-ray binaries, the infrared lags will generally be in the low $\tau$ regime, while from
 \cite{Tetarenko2019,Tetarenko2021}, we can see that the radio bands will generally be in the large $\tau$ regime, and the millimeter bands in the intermediate regime where extra-strong variability could be expected.   For AGN with data taken in a single snapshot, this kind of behavior would be most likely to manifest itself as a spectral energy distribution which deviates more strongly from a single power law than do the typical spectral energy distributions of AGN in the synchrotron-dominated regime.

\begin{figure*}
    \includegraphics[width=3.5 in]{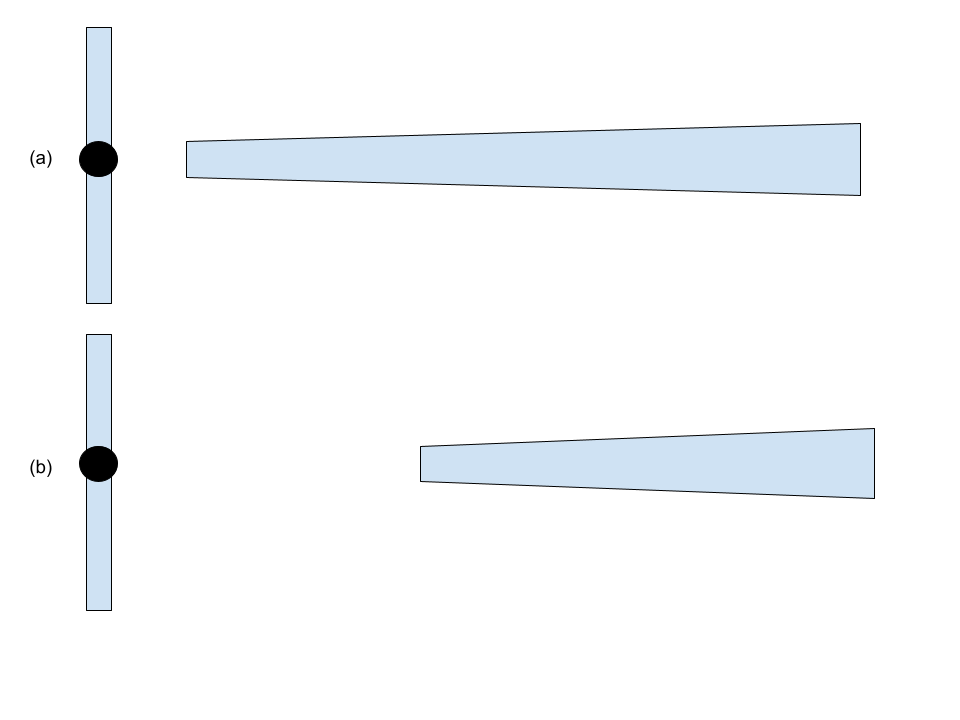} \includegraphics[width=3.5 in]{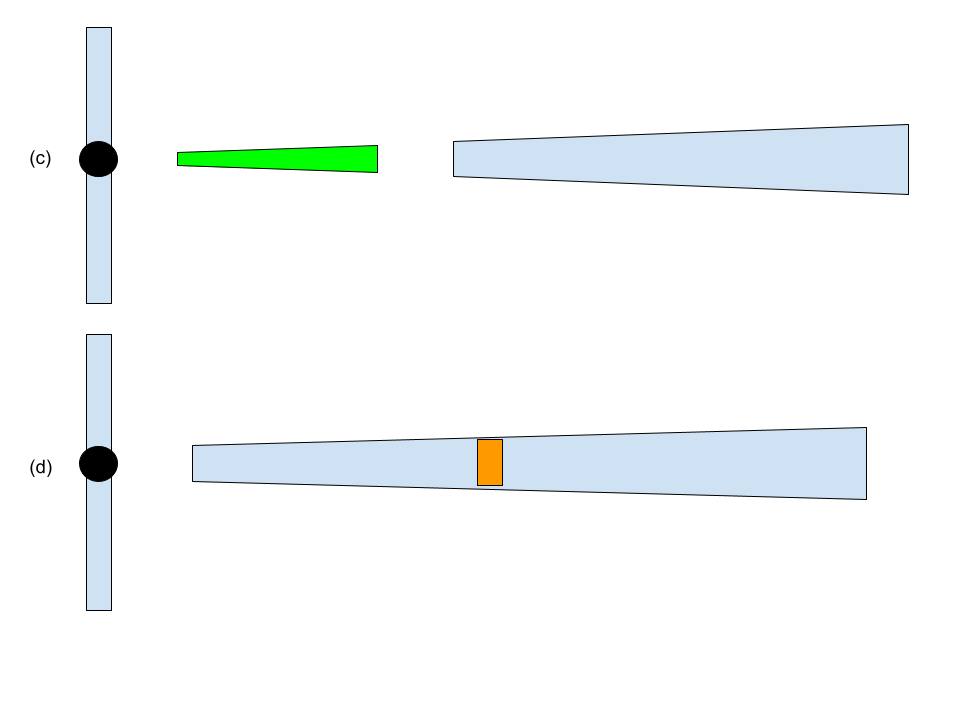}
    \caption{A cartoon illustrating the evolution of the jets in response to the state transitions.  Phase (a) represents the system after it has been in a hard state for an extended period of time, and established its quasi-equilibrium configuration with stochastic variability due to internal shocks.  Phase (b) represents the system after it has settled into a soft state, so that the part of the jet far from the black hole is largely unchanged, but the part near the black hole is missing.  Phase (c) represents the system shortly after it re-enters a hard state, so that the new jet (in green) has started to form and has internal shocking taking place, but it has not yet re-connected with the old jet.  Phase (d) represents the system shortly after the new jet reconnects with the old jet, with the orange region representing the location where the excess energy of the new jet is suddenly dissipated.}
    \label{jetfigure}
\end{figure*}

 { In figure \ref{jetfigure} we show a cartoon schematic of how the jet can be expected to behave through these transitions.  If the system stays in the hard state for an extended period of time, it should establish a quasi-equilibrium in which its jet properties look like a steady hard state, which we illustrate in panel (a).  When the transition happens, energy (and mass) injection into the inner region of the jet ceases, but the outer jet, fueled by material ejected much earlier, will show no effects of the change, which we illustrate in panel (b).  Next, in panel (c) the jet turns back on in the inner region, and the innermost and outermost parts of the jet are powered as in a steady hard state, but the intermediate region is not powered.  Next, in panel (d), the two regions connect, and there is a stronger shock where they meet, as a larger-than-typical amount of power is being dissipated at that junction. 

We can anticipate that in the phase corresponding to panel (a), that the jet will have its standard spectrum and brightness level for a bright hard state.  In the phase corresponding to panel (b), the jet will gradually see its high frequency emission disappear, on a timescale similar to the delay time seen for propagation along the jet to the region where a particular wavelength of light is emitted.  From \cite{Tetarenko2021}, for bright, hard-state X-ray binaries, this will tend to be about $1500 {\rm~sec} (\nu/5~{\rm GHz})^{-1}$.  

Following \cite{2003MNRAS.343L..59H}, the break frequency at which synchrotron emission goes from optically thin to optically thick should scale with $M^{-1/3}$ under the assumptions that all the energetics of the jet behave independently of mass, and only the radiative transfer varies with mass.  Given that the base height, where the jet starts having shocks energize particles will scale linearly with mass, we can expect that the time lags should then scale with mass and frequency as $1500 {\rm~sec} (\nu/5~{\rm GHz})^{-1} (M/10 M_\odot)^{2/3}$.  Characteristic break frequencies are expected to be $\sim10^{14}$ Hz in X-ray binaries and hence $10^{11-12}$ Hz for AGN in the $10^6-10^9 M_\odot$ range \citep{2010MNRAS.404L..21C, 2011ApJ...740L..13G}.  

First, we can consider the case where we ignore the shock in panel (d), and focus on the expectations for phases (b) and (c).  In an X-ray binary where the turnoff time for the jet is about 100 seconds, we can expect to see phenomenology in the jets related to the flip-flop transitions for about 100 seconds, at frequencies above about 80 GHz.  Actually measuring this would require strictly simultaneous data in either the millimeter or infrared band to go with the X-ray data to observe the flip-flops.  Verifying that the jet spectrum corresponds to something like phase (c) at some point would require having at least 3 bands of long wavelength data, in addition to the X-rays, taken simultaneously.  Executing this strategy would be extremely challenging to do by design, because the timing of the flip-flop transitions is unpredictable.  If we take an AGN with a transition timescale of 3 years, and a black hole mass of $10^7 M_\odot$, we find that the variations should be observable from about 35 GHz to the break frequency of $10^{12}$ Hz, precisely in the range of frequencies where CMB experiments provide wide-field monitoring data.  Here, the phenomenon may even, in some cases, be observable first via jet variability.

Second, we can consider whether the shock in phase (d) can be detectable using the constraints from \citet{2019MNRAS.489.4836F}. 
Their equation (29) is:
\begin{equation}
    E_m = 1.5\times 10^{35} {\rm erg} D_{kpc}^{40/17} F^{20/17}_{\nu\rm ,mJy}\nu_{\rm GHz}^{-23/34}
\end{equation}
which applies for optically thick regions of the spectrum.  

For this to reach about 5 mJy at 5~kpc, so that it might dominate over the steadier jet emission, we need about $10^{37}$ erg to be injected.  This is an order of magnitude smaller than the kinetic power put into a jet launched by a $10^{37}$ erg/sec accretion disc, with 10\% jet powering efficiency, over 100 seconds.  It is thus plausible that centimeter-band flare of order a few minutes in duration could be seen if one observed at the right time, but it may be challenging to distinguish from the stochastic jet variability that is often seen \citep{Tetarenko2019}.  

If we move the distance to 100 Mpc, the energy content needed increases by a factor of about $10^{10}$.  We can expect energetics to scale up with $M^2$ (one power for the accretion rate if the Eddington fraction is fixed, and one power for the timescale), so we can then expect that even for $10^6 M_\odot$ black holes, such flaring would be potentially observable as a slow, optically thick transient event. This, too, is consistent with the sources seen in \citet{Nyland2020}, which show very strong radio variability on decade timescales, and most of which show radio spectra which are inverted below about 10 GHz in their rest frames.
}

\subsubsection{Searches for changing state AGN as slow radio transients}

One also might expect to see transient jet behavior in CSAGN.  In fact, the transient AGN seen from the Very Large Array Sky Survey  \citep{Nyland2020} may be examples of this phenomenology, and interestingly, all the systems with good estimates of the Eddington fraction in that sample show Eddington fractions close to 10\%.  In the future, it may be that the best manner to discover these systems is through flaring seen by millimeter-band telescopes  \citep{Guns,SOtransients}{; this would be the case both because the millimeter-band can be expected to have high-duty cycle coverage of a very wide field of view, and because the millimeter band should show more rapid variability than the longer wavelength radio band, as the emission should come from smaller regions in the jet, closer to the central black hole \citep{1979ApJ...232...34B,1985ApJ...298..114M}, something well-established to be the case for the X-ray binaries \citep{Tetarenko2021} and for blazars \citep{2025A&A...702A.140M} and also more tentatively seen in the small fraction of systems with good radio monitoring that are unbeamed AGN \citep{2007A&A...469..899H, 2017ApJ...834..157P}.  Additionally, as noted above, the region of the spectrum where the jet variability is likely to be strongest is exactly the one probed by these surveys}.  Follow-up with multi-epoch spectroscopy of these systems would be warranted.   Similarly, radio follow-up of the optically-selected CSAGN may be fruitful, and at least one example exists of strong radio flaring correlated with a CSAGN  \citep{MeyerLaunch}.

\subsection{Searches for rapid cycling of disc winds}
One of the possible analogous behaviors between X-ray binary and AGN states that has not received much attention is the phenomenology of their disc winds.  Radio-loud AGN are far more likely to show double-peaked emission lines than are radio-quiet AGN  \citep{Eracleous2003}, and this has been explained via the absence of strong winds in the radio-loud systems.  In the X-ray binaries, strong disc winds are seen primarily in soft states, mainly via X-ray absorption lines  \citep{Neilsen2009}, but they can also be inferred from dipping  \citep{DiazTrigo2006} and scattering effects  \citep{Begelman1983}.  

The short durations of the faint states in the flip-flop systems, combined with the fact that they are most frequently observed with relatively low spectral resolution data, mean that searches for rapid turn-off and turn-on of disc winds in these systems has not been done (excepting the case of GRS~1915$+$105  \citealt{Neilsen2009}, which shows some rapid state changes with a different set of characteristics than the flip-flop systems).  In AGN, while this may require relatively long gratings or calorimeter observations { (with facilities like Chandra or XRISM)}, it is straightforward to schedule searches for absorption lines in both states{, and future facilities like NewAthena could be significantly more powerful in such searches \citep{2023hxsi.book..209G}}.

\section{Discussion}

Previous attempts to draw analogies between CSAGN and X-ray binaries have focused on the ``standard" state transitions, which tend to happen on month timescales in X-ray binaries, and hence would be unobservable in AGN if their timescales scaled linearly with black hole mass -- see Table \ref{table_timescales}.  Here we have shown that by associating the CSAGN instead with flip-flop states, we can reconcile the timescales for the changing-state effect very well. While direct evidence linking the two phenomena remains limited, we discuss numerous new observations that could provide stronger connections.  Attacking the problem using both sets of data could help substantially.  Furthermore, the finding that both stellar mass black holes and AGN show essentially the same phenomenology for rapid spectral transitions argues against models in which a mechanism is invoked that is particular to one class of system (e.g. through tidal disruptions --  \citealt{2024ApJ...962L...7W} or through the radiation pressure instability as mentioned in  \citealt{Buisson2025}).

A plausible mechanism for making the timescales so fast is to invoke the thermal timescale.  Near the critical accretion rate for the state transition, gas may be thermally unstable in the accretion disc  \citep{Takizawa1997}.  The thermal timescale for accretion discs is typically $t_{th} = \frac{1}{\alpha \Omega}$.  Thus, if $\alpha$ is similar for AGN and X-ray binaries, the timescale for changes should scale approximately linearly with black hole mass.  

There are two potential strong tests of this picture.  One is that CSAGN are excellent candidates for showing QPOs, with characteristic timescales of a few days to a few months.  Monitoring observations of these systems should be done with sufficient sensitivity and cadence to search for these oscillations.  The second is that in both classes of systems, the jet behavior should be anomalous.   Understanding this may be easier with radio or millimeter-band searches for transient AGN than in X-ray binaries { as the X-ray binary phenomena happen very quickly and unpredictably, meaning one would need to be lucky about when multi-wavelength data were scheduled}.

\section{Acknowledgments}
Several of us thank Gullo Mastroserio for organizing a workshop in Milan where the seeds for this work were planted.  TJM thanks Vanderbilt University, and especially Manuel Pichardo Marcano, for hospitality during a visit in which the idea truly jelled, { and Megan Masterson for useful discussions about AGN QPOs and Doug Lin for discussions about the possibility that CSAGN are driven by tidal disruptions}. GM acknowledges financial support from the Academy of Finland grant 355672.  { This research was supported in part by grant NSF PHY-2309135 to the Kavli Institute for Theoretical Physics (KITP).}  { We thank both the anonymous referee and the editor for comments which have improved the quality of this manuscript.}

\bibliographystyle{plainnat}
\bibliography{bibiliography}

@ARTICLE{2018ApJ...854..160R,
       author = {{Rumbaugh}, N. and {Shen}, Yue and {Morganson}, Eric and {Liu}, Xin and {Banerji}, M. and {McMahon}, R.~G. and {Abdalla}, F.~B. and {Benoit-L{\'e}vy}, A. and {Bertin}, E. and {Brooks}, D. and {Buckley-Geer}, E. and {Capozzi}, D. and {Carnero Rosell}, A. and {Carrasco Kind}, M. and {Carretero}, J. and {Cunha}, C.~E. and {D'Andrea}, C.~B. and {da Costa}, L.~N. and {DePoy}, D.~L. and {Desai}, S. and {Doel}, P. and {Frieman}, J. and {Garc{\'\i}a-Bellido}, J. and {Gruen}, D. and {Gruendl}, R.~A. and {Gschwend}, J. and {Gutierrez}, G. and {Honscheid}, K. and {James}, D.~J. and {Kuehn}, K. and {Kuhlmann}, S. and {Kuropatkin}, N. and {Lima}, M. and {Maia}, M.~A.~G. and {Marshall}, J.~L. and {Martini}, P. and {Menanteau}, F. and {Plazas}, A.~A. and {Reil}, K. and {Roodman}, A. and {Sanchez}, E. and {Scarpine}, V. and {Schindler}, R. and {Schubnell}, M. and {Sheldon}, E. and {Smith}, M. and {Soares-Santos}, M. and {Sobreira}, F. and {Suchyta}, E. and {Swanson}, M.~E.~C. and {Walker}, A.~R. and {Wester}, W. and {DES Collaboration}},
        title = "{Extreme Variability Quasars from the Sloan Digital Sky Survey and the Dark Energy Survey}",
      journal = {\apj},
         year = 2018,
        month = feb,
       volume = {854},
       number = {2},
          eid = {160},
        pages = {160},
          doi = {10.3847/1538-4357/aaa9b6},
archivePrefix = {arXiv},
       eprint = {1706.07875},
 primaryClass = {astro-ph.GA},
       adsurl = {https://ui.adsabs.harvard.edu/abs/2018ApJ...854..160R}
}

@ARTICLE{vanderklis1994,
       author = {{van der Klis}, M.},
        title = "{Similarities in Neutron Star and Black Hole Accretion}",
      journal = {\apjs},
         year = 1994,
        month = jun,
       volume = {92},
        pages = {511},
          doi = {10.1086/192006},
       adsurl = {https://ui.adsabs.harvard.edu/abs/1994ApJS...92..511V}
}

@ARTICLE{edelson2019,
       author = {{Edelson}, R. and {Gelbord}, J. and {Cackett}, E. and {Peterson}, B.~M. and {Horne}, K. and {Barth}, A.~J. and {Starkey}, D.~A. and {Bentz}, M. and {Brandt}, W.~N. and {Goad}, M. and {Joner}, M. and {Korista}, K. and {Netzer}, H. and {Page}, K. and {Uttley}, P. and {Vaughan}, S. and {Breeveld}, A. and {Cenko}, S.~B. and {Done}, C. and {Evans}, P. and {Fausnaugh}, M. and {Ferland}, G. and {Gonzalez-Buitrago}, D. and {Gropp}, J. and {Grupe}, D. and {Kaastra}, J. and {Kennea}, J. and {Kriss}, G. and {Mathur}, S. and {Mehdipour}, M. and {Mudd}, D. and {Nousek}, J. and {Schmidt}, T. and {Vestergaard}, M. and {Villforth}, C.},
        title = "{The First Swift Intensive AGN Accretion Disk Reverberation Mapping Survey}",
      journal = {\apj},
         year = 2019,
        month = jan,
       volume = {870},
       number = {2},
          eid = {123},
        pages = {123},
          doi = {10.3847/1538-4357/aaf3b4},
archivePrefix = {arXiv},
       eprint = {1811.07956},
 primaryClass = {astro-ph.HE},
       adsurl = {https://ui.adsabs.harvard.edu/abs/2019ApJ...870..123E}
}

@ARTICLE{cackett2021,
       author = {{Cackett}, Edward M. and {Bentz}, Misty C. and {Kara}, Erin},
        title = "{Reverberation mapping of active galactic nuclei: from X-ray corona to dusty torus}",
      journal = {iScience},
         year = 2021,
        month = jun,
       volume = {24},
       number = {6},
        pages = {102557},
          doi = {10.1016/j.isci.2021.102557},
archivePrefix = {arXiv},
       eprint = {2105.06926},
 primaryClass = {astro-ph.GA},
       adsurl = {https://ui.adsabs.harvard.edu/abs/2021iSci...24j2557C}
}

@ARTICLE{2024SerAJ.209....1K,
       author = {{Komossa}, S. and {Grupe}, D.},
        title = "{The Extremes of Continuum and Emission-Line Variability of AGN: Changing-Look Events and Binary SMBHS}",
      journal = {Serbian Astronomical Journal},
         year = 2024,
        month = dec,
       volume = {209},
        pages = {1-24},
          doi = {10.2298/SAJ2409001K},
       adsurl = {https://ui.adsabs.harvard.edu/abs/2024SerAJ.209....1K}
}

@ARTICLE{2004A&A...426..587C,
       author = {{Casella}, P. and {Belloni}, T. and {Homan}, J. and {Stella}, L.},
        title = "{A study of the low-frequency quasi-periodic oscillations in the X-ray light curves of the black hole candidate <ASTROBJ>XTE J1859+226</ASTROBJ>}",
      journal = {\aap},
         year = 2004,
        month = nov,
       volume = {426},
        pages = {587-600},
          doi = {10.1051/0004-6361:20041231},
archivePrefix = {arXiv},
       eprint = {astro-ph/0407262},
 primaryClass = {astro-ph},
       adsurl = {https://ui.adsabs.harvard.edu/abs/2004A&A...426..587C}
}

@INCOLLECTION{2023hxsi.book..209G,
       author = {{Gallo}, Luigi C. and {Miller}, Jon M. and {Costantini}, Elisa},
        title = "{Active Galactic Nuclei with High-Resolution X-Ray Spectroscopy}",
    booktitle = {High-Resolution X-ray Spectroscopy: Instrumentation},
         year = 2023,
       editor = {{Bambi}, Cosimo and {Jiang}, Jiachen},
        pages = {209-254},
          doi = {10.1007/978-981-99-4409-5_9},
       adsurl = {https://ui.adsabs.harvard.edu/abs/2023hxsi.book..209G}
}

@ARTICLE{1979ApJ...232...34B,
       author = {{Blandford}, R.~D. and {K{\"o}nigl}, A.},
        title = "{Relativistic jets as compact radio sources.}",
      journal = {\apj},
         year = 1979,
        month = aug,
       volume = {232},
        pages = {34-48},
          doi = {10.1086/157262},
       adsurl = {https://ui.adsabs.harvard.edu/abs/1979ApJ...232...34B}
}

@ARTICLE{2025A&A...702A.140M,
       author = {{Marchili}, N. and {Righini}, S. and {Giroletti}, M. and {Raiteri}, C.~M. and {Giri}, R.~P. and {Carnerero}, M.~I. and {Villata}, M. and {Bach}, U. and {Cassaro}, P. and {Liuzzo}, E. and {Buemi}, C.~S. and {Leto}, P. and {Trigilio}, C. and {Umana}, G. and {Bonato}, M. and {Patricelli}, B. and {Stamerra}, A.},
        title = "{Twenty years of blazar monitoring with the INAF radio telescopes}",
      journal = {\aap},
         year = 2025,
        month = oct,
       volume = {702},
          eid = {A140},
        pages = {A140},
          doi = {10.1051/0004-6361/202556630},
archivePrefix = {arXiv},
       eprint = {2509.04577},
 primaryClass = {astro-ph.HE},
       adsurl = {https://ui.adsabs.harvard.edu/abs/2025A&A...702A.140M}
}

@ARTICLE{1985ApJ...298..114M,
       author = {{Marscher}, A.~P. and {Gear}, W.~K.},
        title = "{Models for high-frequency radio outbursts in extragalactic sources, with application to the early 1983 millimeter-to-infrared flare of 3C 273.}",
      journal = {\apj},
         year = 1985,
        month = nov,
       volume = {298},
        pages = {114-127},
          doi = {10.1086/163592},
       adsurl = {https://ui.adsabs.harvard.edu/abs/1985ApJ...298..114M}
}

@ARTICLE{2007A&A...469..899H,
       author = {{Hovatta}, T. and {Tornikoski}, M. and {Lainela}, M. and {Lehto}, H.~J. and {Valtaoja}, E. and {Torniainen}, I. and {Aller}, M.~F. and {Aller}, H.~D.},
        title = "{Statistical analyses of long-term variability of AGN at high radio frequencies}",
      journal = {\aap},
         year = 2007,
        month = jul,
       volume = {469},
       number = {3},
        pages = {899-912},
          doi = {10.1051/0004-6361:20077529},
archivePrefix = {arXiv},
       eprint = {0705.3293},
 primaryClass = {astro-ph},
       adsurl = {https://ui.adsabs.harvard.edu/abs/2007A&A...469..899H}
}

@ARTICLE{2017ApJ...834..157P,
       author = {{Park}, Jongho and {Trippe}, Sascha},
        title = "{The Long-term Centimeter Variability of Active Galactic Nuclei: A New Relation between Variability Timescale and Accretion Rate}",
      journal = {\apj},
         year = 2017,
        month = jan,
       volume = {834},
       number = {2},
          eid = {157},
        pages = {157},
          doi = {10.3847/1538-4357/834/2/157},
archivePrefix = {arXiv},
       eprint = {1611.04729},
 primaryClass = {astro-ph.HE},
       adsurl = {https://ui.adsabs.harvard.edu/abs/2017ApJ...834..157P}
}

@ARTICLE{2000ARA&A..38..717V,
       author = {{van der Klis}, M.},
        title = "{Millisecond Oscillations in X-ray Binaries}",
      journal = {\araa},
         year = 2000,
        month = jan,
       volume = {38},
        pages = {717-760},
          doi = {10.1146/annurev.astro.38.1.717},
archivePrefix = {arXiv},
       eprint = {astro-ph/0001167},
 primaryClass = {astro-ph},
       adsurl = {https://ui.adsabs.harvard.edu/abs/2000ARA&A..38..717V}
}

@ARTICLE{2003MNRAS.343L..59H,
       author = {{Heinz}, S. and {Sunyaev}, R.~A.},
        title = "{The non-linear dependence of flux on black hole mass and accretion rate in core-dominated jets}",
      journal = {\mnras},
         year = 2003,
        month = aug,
       volume = {343},
       number = {3},
        pages = {L59-L64},
          doi = {10.1046/j.1365-8711.2003.06918.x},
archivePrefix = {arXiv},
       eprint = {astro-ph/0305252},
 primaryClass = {astro-ph},
       adsurl = {https://ui.adsabs.harvard.edu/abs/2003MNRAS.343L..59H}
}

@ARTICLE{2011ApJ...740L..13G,
       author = {{Gandhi}, P. and {Blain}, A.~W. and {Russell}, D.~M. and {Casella}, P. and {Malzac}, J. and {Corbel}, S. and {D'Avanzo}, P. and {Lewis}, F.~W. and {Markoff}, S. and {Cadolle Bel}, M. and {Goldoni}, P. and {Wachter}, S. and {Khangulyan}, D. and {Mainzer}, A.},
        title = "{A Variable Mid-infrared Synchrotron Break Associated with the Compact Jet in GX 339-4}",
      journal = {\apjl},
         year = 2011,
        month = oct,
       volume = {740},
       number = {1},
          eid = {L13},
        pages = {L13},
          doi = {10.1088/2041-8205/740/1/L13},
archivePrefix = {arXiv},
       eprint = {1109.4143},
 primaryClass = {astro-ph.HE},
       adsurl = {https://ui.adsabs.harvard.edu/abs/2011ApJ...740L..13G}
}

@ARTICLE{2019MNRAS.489.4836F,
       author = {{Fender}, Rob and {Bright}, Joe},
        title = "{Synchrotron self-absorption and the minimum energy of optically thick radio flares from stellar mass black holes}",
      journal = {\mnras},
         year = 2019,
        month = nov,
       volume = {489},
       number = {4},
        pages = {4836-4846},
          doi = {10.1093/mnras/stz2000},
archivePrefix = {arXiv},
       eprint = {1907.07463},
 primaryClass = {astro-ph.HE},
       adsurl = {https://ui.adsabs.harvard.edu/abs/2019MNRAS.489.4836F}
}

@ARTICLE{2018MNRAS.480.3898N,
       author = {{Noda}, Hirofumi and {Done}, Chris},
        title = "{Explaining changing-look AGN with state transition triggered by rapid mass accretion rate drop}",
      journal = {\mnras},
         year = 2018,
        month = nov,
       volume = {480},
       number = {3},
        pages = {3898-3906},
          doi = {10.1093/mnras/sty2032},
archivePrefix = {arXiv},
       eprint = {1805.07873},
 primaryClass = {astro-ph.GA},
       adsurl = {https://ui.adsabs.harvard.edu/abs/2018MNRAS.480.3898N}
}

@ARTICLE{2019RNAAS...3..116M,
       author = {{Maccarone}, Thomas J.},
        title = "{Detection of Variability Features in Background Limited Data}",
      journal = {Research Notes of the American Astronomical Society},
         year = 2019,
        month = aug,
       volume = {3},
       number = {8},
          eid = {116},
        pages = {116},
          doi = {10.3847/2515-5172/ab3a36},
       adsurl = {https://ui.adsabs.harvard.edu/abs/2019RNAAS...3..116M}
}

@ARTICLE{2016MNRAS.461.3145V,
       author = {{Vaughan}, S. and {Uttley}, P. and {Markowitz}, A.~G. and {Huppenkothen}, D. and {Middleton}, M.~J. and {Alston}, W.~N. and {Scargle}, J.~D. and {Farr}, W.~M.},
        title = "{False periodicities in quasar time-domain surveys}",
      journal = {\mnras},
         year = 2016,
        month = sep,
       volume = {461},
       number = {3},
        pages = {3145-3152},
          doi = {10.1093/mnras/stw1412},
archivePrefix = {arXiv},
       eprint = {1606.02620},
 primaryClass = {astro-ph.IM},
       adsurl = {https://ui.adsabs.harvard.edu/abs/2016MNRAS.461.3145V}
}

@ARTICLE{2025ApJ...986..160D,
       author = {{Dong}, Qian and {Zhang}, Zhi-Xiang and {Gu}, Wei-Min and {Sun}, Mouyuan and {Zheng}, Yong-Gang},
        title = "{Newly Discovered Changing-look Active Galactic Nuclei from SDSS and LAMOST Survey}",
      journal = {\apj},
         year = 2025,
        month = jun,
       volume = {986},
       number = {2},
          eid = {160},
        pages = {160},
          doi = {10.3847/1538-4357/add331},
archivePrefix = {arXiv},
       eprint = {2408.07335},
 primaryClass = {astro-ph.GA},
       adsurl = {https://ui.adsabs.harvard.edu/abs/2025ApJ...986..160D}
}

@ARTICLE{MeyerLaunch,
       author = {{Meyer}, Eileen T. and {Laha}, Sibasish and {Shuvo}, Onic I. and {Roychowdhury}, Agniva and {Green}, David A. and {Rhodes}, Lauren and {Hankla}, Amelia M. and {Philippov}, Alexander and {Mbarek}, Rostom and {laor}, Ari and {Begelman}, Mitchell C. and {Sadaula}, Dev R. and {Ghosh}, Ritesh and {Bruni}, Gabriele and {Panessa}, Francesca and {Guainazzi}, Matteo and {Behar}, Ehud and {Masterson}, Megan and {Zhang}, Haocheng and {Yang}, Xiaolong and {Gurwell}, Mark A. and {Keating}, Garrett K. and {Williams-Baldwin}, David and {Bray}, Justin D. and {Bempong-Manful}, Emmanuel K. and {Wrigley}, Nicholas and {Bianchi}, Stefano and {Ricci}, Federica and {La Franca}, Fabio and {Kara}, Erin and {Georganopoulos}, Markos and {Oates}, Samantha and {Nicholl}, Matt and {Pal}, Main and {Cenko}, S. Bradley},
        title = "{Late-time Radio Brightening and Emergence of a Radio Jet in the Changing-look AGN 1ES 1927+654}",
      journal = {\apjl},
         year = 2025,
        month = jan,
       volume = {979},
       number = {1},
          eid = {L2},
        pages = {L2},
          doi = {10.3847/2041-8213/ad8651},
archivePrefix = {arXiv},
       eprint = {2406.18061},
 primaryClass = {astro-ph.HE},
       adsurl = {https://ui.adsabs.harvard.edu/abs/2025ApJ...979L...2M}
}

@ARTICLE{Ingram2009,
       author = {{Ingram}, Adam and {Done}, Chris and {Fragile}, P. Chris},
        title = "{Low-frequency quasi-periodic oscillations spectra and Lense-Thirring precession}",
      journal = {\mnras},
         year = 2009,
        month = jul,
       volume = {397},
       number = {1},
        pages = {L101-L105},
          doi = {10.1111/j.1745-3933.2009.00693.x},
archivePrefix = {arXiv},
       eprint = {0901.1238},
 primaryClass = {astro-ph.SR},
       adsurl = {https://ui.adsabs.harvard.edu/abs/2009MNRAS.397L.101I}
}

@ARTICLE{BelloniQPOhardness,
       author = {{Belloni}, T. and {van der Klis}, M. and {Lewin}, W.~H.~G. and {van Paradijs}, J. and {Dotani}, T. and {Mitsuda}, K. and {Miyamoto}, S.},
        title = "{Energy dependence in the quasi-periodic oscillations and noise of black hole candidates in the very high state.}",
      journal = {\aap},
         year = 1997,
        month = jun,
       volume = {322},
        pages = {857-867},
       adsurl = {https://ui.adsabs.harvard.edu/abs/1997A&A...322..857B}
}

@ARTICLE{Dunn2025,
       author = {{Dunn}, Thomas and {McElroy}, Rebecca and {Krumpe}, Mirko and {Croom}, Scott M. and {Gaspari}, Massimo and {Perez-Torres}, Miguel and {Cowley}, Michael and {Omoruyi}, Osase and {Tremblay}, Grant and {Singha}, Mainak},
        title = "{A century of change: new changing-look event in Mrk 1018's past}",
      journal = {arXiv e-prints},
         year = 2025,
        month = oct,
          eid = {arXiv:2510.25156},
        pages = {arXiv:2510.25156},
          doi = {10.48550/arXiv.2510.25156},
archivePrefix = {arXiv},
       eprint = {2510.25156},
 primaryClass = {astro-ph.GA},
       adsurl = {https://ui.adsabs.harvard.edu/abs/2025arXiv251025156D}
}

@ARTICLE{2024ApJ...962L...7W,
       author = {{Wang}, Yihan and {Lin}, Douglas N.~C. and {Zhang}, Bing and {Zhu}, Zhaohuan},
        title = "{Changing-look Active Galactic Nuclei Behavior Induced by Disk-captured Tidal Disruption Events}",
      journal = {\apjl},
         year = 2024,
        month = feb,
       volume = {962},
       number = {1},
          eid = {L7},
        pages = {L7},
          doi = {10.3847/2041-8213/ad20e5},
archivePrefix = {arXiv},
       eprint = {2310.00038},
 primaryClass = {astro-ph.HE},
       adsurl = {https://ui.adsabs.harvard.edu/abs/2024ApJ...962L...7W}
}

@ARTICLE{1998A&A...337..460H,
       author = {{Hannikainen}, D.~C. and {Hunstead}, R.~W. and {Campbell-Wilson}, D. and {Sood}, R.~K.},
        title = "{MOST radio monitoring of GX 339-4}",
      journal = {\aap},
         year = 1998,
        month = sep,
       volume = {337},
        pages = {460-464},
          doi = {10.48550/arXiv.astro-ph/9805332},
archivePrefix = {arXiv},
       eprint = {astro-ph/9805332},
 primaryClass = {astro-ph},
       adsurl = {https://ui.adsabs.harvard.edu/abs/1998A&A...337..460H}
}

@ARTICLE{LSST,
       author = {{Ivezi{\'c}}, {\v{Z}}eljko and {Kahn}, Steven M. and {Tyson}, J. Anthony and {Abel}, Bob and {Acosta}, Emily and {Allsman}, Robyn and {Alonso}, David and {AlSayyad}, Yusra and {Anderson}, Scott F. and {Andrew}, John and {Angel}, James Roger P. and {Angeli}, George Z. and {Ansari}, Reza and {Antilogus}, Pierre and {Araujo}, Constanza and {Armstrong}, Robert and {Arndt}, Kirk T. and {Astier}, Pierre and {Aubourg}, {\'E}ric and {Auza}, Nicole and {Axelrod}, Tim S. and {Bard}, Deborah J. and {Barr}, Jeff D. and {Barrau}, Aurelian and {Bartlett}, James G. and {Bauer}, Amanda E. and {Bauman}, Brian J. and {Baumont}, Sylvain and {Bechtol}, Ellen and {Bechtol}, Keith and {Becker}, Andrew C. and {Becla}, Jacek and {Beldica}, Cristina and {Bellavia}, Steve and {Bianco}, Federica B. and {Biswas}, Rahul and {Blanc}, Guillaume and {Blazek}, Jonathan and {Blandford}, Roger D. and {Bloom}, Josh S. and {Bogart}, Joanne and {Bond}, Tim W. and {Booth}, Michael T. and {Borgland}, Anders W. and {Borne}, Kirk and {Bosch}, James F. and {Boutigny}, Dominique and {Brackett}, Craig A. and {Bradshaw}, Andrew and {Brandt}, William Nielsen and {Brown}, Michael E. and {Bullock}, James S. and {Burchat}, Patricia and {Burke}, David L. and {Cagnoli}, Gianpietro and {Calabrese}, Daniel and {Callahan}, Shawn and {Callen}, Alice L. and {Carlin}, Jeffrey L. and {Carlson}, Erin L. and {Chandrasekharan}, Srinivasan and {Charles-Emerson}, Glenaver and {Chesley}, Steve and {Cheu}, Elliott C. and {Chiang}, Hsin-Fang and {Chiang}, James and {Chirino}, Carol and {Chow}, Derek and {Ciardi}, David R. and {Claver}, Charles F. and {Cohen-Tanugi}, Johann and {Cockrum}, Joseph J. and {Coles}, Rebecca and {Connolly}, Andrew J. and {Cook}, Kem H. and {Cooray}, Asantha and {Covey}, Kevin R. and {Cribbs}, Chris and {Cui}, Wei and {Cutri}, Roc and {Daly}, Philip N. and {Daniel}, Scott F. and {Daruich}, Felipe and {Daubard}, Guillaume and {Daues}, Greg and {Dawson}, William and {Delgado}, Francisco and {Dellapenna}, Alfred and {de Peyster}, Robert and {de Val-Borro}, Miguel and {Digel}, Seth W. and {Doherty}, Peter and {Dubois}, Richard and {Dubois-Felsmann}, Gregory P. and {Durech}, Josef and {Economou}, Frossie and {Eifler}, Tim and {Eracleous}, Michael and {Emmons}, Benjamin L. and {Fausti Neto}, Angelo and {Ferguson}, Henry and {Figueroa}, Enrique and {Fisher-Levine}, Merlin and {Focke}, Warren and {Foss}, Michael D. and {Frank}, James and {Freemon}, Michael D. and {Gangler}, Emmanuel and {Gawiser}, Eric and {Geary}, John C. and {Gee}, Perry and {Geha}, Marla and {Gessner}, Charles J.~B. and {Gibson}, Robert R. and {Gilmore}, D. Kirk and {Glanzman}, Thomas and {Glick}, William and {Goldina}, Tatiana and {Goldstein}, Daniel A. and {Goodenow}, Iain and {Graham}, Melissa L. and {Gressler}, William J. and {Gris}, Philippe and {Guy}, Leanne P. and {Guyonnet}, Augustin and {Haller}, Gunther and {Harris}, Ron and {Hascall}, Patrick A. and {Haupt}, Justine and {Hernandez}, Fabio and {Herrmann}, Sven and {Hileman}, Edward and {Hoblitt}, Joshua and {Hodgson}, John A. and {Hogan}, Craig and {Howard}, James D. and {Huang}, Dajun and {Huffer}, Michael E. and {Ingraham}, Patrick and {Innes}, Walter R. and {Jacoby}, Suzanne H. and {Jain}, Bhuvnesh and {Jammes}, Fabrice and {Jee}, M. James and {Jenness}, Tim and {Jernigan}, Garrett and {Jevremovi{\'c}}, Darko and {Johns}, Kenneth and {Johnson}, Anthony S. and {Johnson}, Margaret W.~G. and {Jones}, R. Lynne and {Juramy-Gilles}, Claire and {Juri{\'c}}, Mario and {Kalirai}, Jason S. and {Kallivayalil}, Nitya J. and {Kalmbach}, Bryce and {Kantor}, Jeffrey P. and {Karst}, Pierre and {Kasliwal}, Mansi M. and {Kelly}, Heather and {Kessler}, Richard and {Kinnison}, Veronica and {Kirkby}, David and {Knox}, Lloyd and {Kotov}, Ivan V. and {Krabbendam}, Victor L. and {Krughoff}, K. Simon and {Kub{\'a}nek}, Petr and {Kuczewski}, John and {Kulkarni}, Shri and {Ku}, John and {Kurita}, Nadine R. and {Lage}, Craig S. and {Lambert}, Ron and {Lange}, Travis and {Langton}, J. Brian and {Le Guillou}, Laurent and {Levine}, Deborah and {Liang}, Ming and {Lim}, Kian-Tat and {Lintott}, Chris J. and {Long}, Kevin E. and {Lopez}, Margaux and {Lotz}, Paul J. and {Lupton}, Robert H. and {Lust}, Nate B. and {MacArthur}, Lauren A. and {Mahabal}, Ashish and {Mandelbaum}, Rachel and {Markiewicz}, Thomas W. and {Marsh}, Darren S. and {Marshall}, Philip J. and {Marshall}, Stuart and {May}, Morgan and {McKercher}, Robert and {McQueen}, Michelle and {Meyers}, Joshua and {Migliore}, Myriam and {Miller}, Michelle and {Mills}, David J.},
        title = "{LSST: From Science Drivers to Reference Design and Anticipated Data Products}",
      journal = {\apj},
         year = 2019,
        month = mar,
       volume = {873},
       number = {2},
          eid = {111},
        pages = {111},
          doi = {10.3847/1538-4357/ab042c},
archivePrefix = {arXiv},
       eprint = {0805.2366},
 primaryClass = {astro-ph},
       adsurl = {https://ui.adsabs.harvard.edu/abs/2019ApJ...873..111I}
}

@INPROCEEDINGS{Argus,
       author = {{Law}, Nicholas and {Corbett}, Hank and {Vasquez Soto}, Alan and {Gonzalez}, Ramses and {Machia}, Lawrence and {Carney}, Jonathan and {Fitton}, Shannon and {Galliher}, Nathan and {Glazier}, Amy and {Marshall}, William and {Procter}, Thomas and {Walters}, Glenn and {Argus Array Science Team}},
        title = "{The Argus Array: Low-cost Access to the Deep, High-Cadence Sky}",
    booktitle = {American Astronomical Society Meeting Abstracts \#243},
         year = 2024,
       series = {American Astronomical Society Meeting Abstracts},
       volume = {243},
        month = feb,
          eid = {237.06},
        pages = {237.06},
       adsurl = {https://ui.adsabs.harvard.edu/abs/2024AAS...24323706L}
}

@ARTICLE{2018ApJ...860L..10S,
       author = {{Smith}, Krista Lynne and {Mushotzky}, Richard F. and {Boyd}, Patricia T. and {Wagoner}, Robert V.},
        title = "{Evidence for an Optical Low-frequency Quasi-periodic Oscillation in the Kepler Light Curve of an Active Galaxy}",
      journal = {\apjl},
         year = 2018,
        month = jun,
       volume = {860},
       number = {1},
          eid = {L10},
        pages = {L10},
          doi = {10.3847/2041-8213/aac88c},
archivePrefix = {arXiv},
       eprint = {1805.12154},
 primaryClass = {astro-ph.HE},
       adsurl = {https://ui.adsabs.harvard.edu/abs/2018ApJ...860L..10S}
}

@ARTICLE{Ray2024,
       author = {{Ray}, Paul S. and {Roming}, Peter W.~A. and {Argan}, Andrea and {Arzoumanian}, Zaven and {Ballantyne}, David R. and {Bogdanov}, Slavko and {Bonvicini}, Valter and {Brandt}, Terri J. and {Bursa}, Michal and {Cackett}, Edward M. and {Chakrabarty}, Deepto and {Christophersen}, Marc and {Coderre}, Kathleen M. and {De Geronimo}, Gianluigi and {Monte}, Ettore Del and {DeRosa}, Alessandra and {Dietz}, Harley R. and {Evangelista}, Yuri and {Feroci}, Marco and {Ford}, Jeremy J. and {Froning}, Cynthia and {Fryer}, Christopher L. and {Gendreau}, Keith C. and {Goldstein}, Adam and {Gonzalez}, Anthony H. and {Hartmann}, Dieter and {Hernanz}, Margarita and {Hutcheson}, Anthony and {in't Zand}, Jean and {Jenke}, Peter and {Kennea}, Jamie and {Lloyd-Ronning}, Nicole M. and {Maccarone}, Thomas J. and {Maes}, Dominic and {Markwardt}, Craig B. and {Michalska}, Malgorzata and {Okajima}, Takashi and {Patruno}, Alessandro and {Persyn}, Steven C. and {Phillips}, Mark L. and {Prescod-Weinstein}, Chanda and {Redfern}, Jillian A. and {Remillard}, Ronald A. and {Santangelo}, Andrea and {Schwendeman}, Carl L. and {Sleator}, Clio and {Steiner}, James and {Strohmayer}, Tod E. and {Svoboda}, Jiri and {Tenzer}, Christoph and {Thompson}, Steven P. and {Warwick}, Richard W. and {Watts}, Anna L. and {Wilson-Hodge}, Colleen A. and {Wu}, Xin and {Wulf}, Eric A. and {Zampa}, Gianluigi},
        title = "{Spectroscopic Time-Resolving Observatory for Broadband Energy X-rays: mission overview}",
      journal = {Journal of Astronomical Telescopes, Instruments, and Systems},
         year = 2024,
        month = oct,
       volume = {10},
          eid = {042504},
        pages = {042504},
          doi = {10.1117/1.JATIS.10.4.042504},
       adsurl = {https://ui.adsabs.harvard.edu/abs/2024JATIS..10d2504R}
}

@INPROCEEDINGS{AXISPaper,
       author = {{Reynolds}, Christopher S. and {Kara}, Erin A. and {Mushotzky}, Richard F. and {Ptak}, Andrew and {Koss}, Michael J. and {Williams}, Brian J. and {Allen}, Steven W. and {Bauer}, Franz E. and {Bautz}, Marshall and {Bogadhee}, Arash and {Burdge}, Kevin B. and {Cappelluti}, Nico and {Cenko}, Brad and {Chartas}, George and {Chan}, Kai-Wing and {Corrales}, L{\'\i}a. and {Daylan}, Tansu and {Falcone}, Abraham D. and {Foord}, Adi and {Grant}, Catherine E. and {Habouzit}, M{\'e}lanie and {Haggard}, Daryl and {Herrmann}, Sven and {Hodges-Kluck}, Edmund and {Kargaltsev}, Oleg and {King}, George W. and {Kounkel}, Marina and {Lopez}, Laura A. and {Marchesi}, Stefano and {McDonald}, Michael and {Meyer}, Eileen and {Miller}, Eric D. and {Nynka}, Melania and {Okajima}, Takashi and {Pacucci}, Fabio and {Russell}, Helen R. and {Safi-Harb}, Samar and {Strassun}, Keivan G. and {Trindade Falc{\~a}o}, Anna and {Walker}, Stephen A. and {Wilms}, Joern and {Yukita}, Mihoko and {Zhang}, William W.},
        title = "{Overview of the advanced x-ray imaging satellite (AXIS)}",
    booktitle = {UV, X-Ray, and Gamma-Ray Space Instrumentation for Astronomy XXIII},
         year = 2023,
       editor = {{Siegmund}, Oswald H. and {Hoadley}, Keri},
       series = {Society of Photo-Optical Instrumentation Engineers (SPIE) Conference Series},
       volume = {12678},
        month = oct,
          eid = {126781E},
        pages = {126781E},
          doi = {10.1117/12.2677468},
archivePrefix = {arXiv},
       eprint = {2311.00780},
 primaryClass = {astro-ph.IM},
       adsurl = {https://ui.adsabs.harvard.edu/abs/2023SPIE12678E..1ER}
}

@ARTICLE{Maccarone2020,
       author = {{Maccarone}, Thomas J. and {Osler}, Arlo and {Miller-Jones}, James C.~A. and {Atri}, P. and {Russell}, David M. and {Meier}, David L. and {McHardy}, Ian M. and {Longa-Pe{\~n}a}, Penelope A.},
        title = "{The stringent upper limit on jet power in the persistent soft-state source 4U 1957+11}",
      journal = {\mnras},
         year = 2020,
        month = oct,
       volume = {498},
       number = {1},
        pages = {L40-L45},
          doi = {10.1093/mnrasl/slaa120},
archivePrefix = {arXiv},
       eprint = {2007.00834},
 primaryClass = {astro-ph.HE},
       adsurl = {https://ui.adsabs.harvard.edu/abs/2020MNRAS.498L..40M}
}

@ARTICLE{Varniere2017,
       author = {{Varniere}, Peggy and {Vincent}, Frederic H.},
        title = "{Reproducing the Correlations of Type C Low-frequency Quasi-periodic Oscillation Parameters in XTE J1550-564 with a Spiral Structure}",
      journal = {\apj},
         year = 2017,
        month = jan,
       volume = {834},
       number = {2},
          eid = {188},
        pages = {188},
          doi = {10.3847/1538-4357/834/2/188},
       adsurl = {https://ui.adsabs.harvard.edu/abs/2017ApJ...834..188V}
}

@ARTICLE{StellaVietri,
       author = {{Stella}, Luigi and {Vietri}, Mario},
        title = "{Lense-Thirring Precession and Quasi-periodic Oscillations in Low-Mass X-Ray Binaries}",
      journal = {\apjl},
         year = 1998,
        month = jan,
       volume = {492},
       number = {1},
        pages = {L59-L62},
          doi = {10.1086/311075},
archivePrefix = {arXiv},
       eprint = {astro-ph/9709085},
 primaryClass = {astro-ph},
       adsurl = {https://ui.adsabs.harvard.edu/abs/1998ApJ...492L..59S}
}

@ARTICLE{Motta2018,
       author = {{Motta}, S.~E. and {Franchini}, A. and {Lodato}, G. and {Mastroserio}, G.},
        title = "{On the different flavours of Lense-Thirring precession around accreting stellar mass black holes}",
      journal = {\mnras},
         year = 2018,
        month = jan,
       volume = {473},
       number = {1},
        pages = {431-439},
          doi = {10.1093/mnras/stx2358},
archivePrefix = {arXiv},
       eprint = {1709.02608},
 primaryClass = {astro-ph.HE},
       adsurl = {https://ui.adsabs.harvard.edu/abs/2018MNRAS.473..431M}
}

@ARTICLE{Motta2015,
       author = {{Motta}, S.~E. and {Casella}, P. and {Henze}, M. and {Mu{\~n}oz-Darias}, T. and {Sanna}, A. and {Fender}, R. and {Belloni}, T.},
        title = "{Geometrical constraints on the origin of timing signals from black holes}",
      journal = {\mnras},
         year = 2015,
        month = feb,
       volume = {447},
       number = {2},
        pages = {2059-2072},
          doi = {10.1093/mnras/stu2579},
archivePrefix = {arXiv},
       eprint = {1404.7293},
 primaryClass = {astro-ph.HE},
       adsurl = {https://ui.adsabs.harvard.edu/abs/2015MNRAS.447.2059M}
}

@ARTICLE{Zhang1997,
       author = {{Zhang}, S.~N. and {Cui}, Wei and {Chen}, Wan},
        title = "{Black Hole Spin in X-Ray Binaries: Observational Consequences}",
      journal = {\apjl},
         year = 1997,
        month = jun,
       volume = {482},
       number = {2},
        pages = {L155-L158},
          doi = {10.1086/310705},
archivePrefix = {arXiv},
       eprint = {astro-ph/9704072},
 primaryClass = {astro-ph},
       adsurl = {https://ui.adsabs.harvard.edu/abs/1997ApJ...482L.155Z}
}

@ARTICLE{DongLAMOSTsample,
       author = {{Dong}, Qian and {Zhang}, Zhi-Xiang and {Gu}, Wei-Min and {Sun}, Mouyuan and {Guo}, Wei-Jian and {Cai}, Zhen-Yi and {Wang}, Jun-Xian and {Zheng}, Yong-Gang},
        title = "{Discovery of Repeating Transitions in 25 Changing-look Active Galactic Nuclei}",
      journal = {arXiv e-prints},
         year = 2025,
        month = oct,
          eid = {arXiv:2510.18445},
        pages = {arXiv:2510.18445},
          doi = {10.48550/arXiv.2510.18445},
archivePrefix = {arXiv},
       eprint = {2510.18445},
 primaryClass = {astro-ph.GA},
       adsurl = {https://ui.adsabs.harvard.edu/abs/2025arXiv251018445D}
}

@ARTICLE{Motch1983,
       author = {{Motch}, C. and {Ricketts}, M.~J. and {Page}, C.~G. and {Ilovaisky}, S.~A. and {Chevalier}, C.},
        title = "{Simultaneous X-ray/optical observations of GX 339-4 during the May 1981 optically bright state.}",
      journal = {\aap},
         year = 1983,
        month = mar,
       volume = {119},
        pages = {171-176},
       adsurl = {https://ui.adsabs.harvard.edu/abs/1983A&A...119..171M}
}

@ARTICLE{Koljonen,
       author = {{Koljonen}, Karri I.~I. and {Satalecka}, Konstancja and {Lindfors}, Elina J. and {Liodakis}, Ioannis},
        title = "{Microquasar Cyg X-3 - a unique jet-wind neutrino factory?}",
      journal = {\mnras},
         year = 2023,
        month = sep,
       volume = {524},
       number = {1},
        pages = {L89-L93},
          doi = {10.1093/mnrasl/slad081},
archivePrefix = {arXiv},
       eprint = {2306.11804},
 primaryClass = {astro-ph.HE},
       adsurl = {https://ui.adsabs.harvard.edu/abs/2023MNRAS.524L..89K}
}

@ARTICLE{HeinzSunyaev,
       author = {{Heinz}, S. and {Sunyaev}, R.},
        title = "{Cosmic rays from microquasars: A narrow component to the CR spectrum?}",
      journal = {\aap},
         year = 2002,
        month = aug,
       volume = {390},
        pages = {751-766},
          doi = {10.1051/0004-6361:20020615},
archivePrefix = {arXiv},
       eprint = {astro-ph/0204183},
 primaryClass = {astro-ph},
       adsurl = {https://ui.adsabs.harvard.edu/abs/2002A&A...390..751H}
}

@ARTICLE{Punch,
       author = {{Punch}, M. and {Akerlof}, C.~W. and {Cawley}, M.~F. and {Chantell}, M. and {Fegan}, D.~J. and {Fennell}, S. and {Gaidos}, J.~A. and {Hagan}, J. and {Hillas}, A.~M. and {Jiang}, Y. and {Kerrick}, A.~D. and {Lamb}, R.~C. and {Lawrence}, M.~A. and {Lewis}, D.~A. and {Meyer}, D.~I. and {Mohanty}, G. and {O'Flaherty}, K.~S. and {Reynolds}, P.~T. and {Rovero}, A.~C. and {Schubnell}, M.~S. and {Sembroski}, G. and {Weekes}, T.~C. and {Whitaker}, T. and {Wilson}, C.},
        title = "{Detection of TeV photons from the active galaxy Markarian 421}",
      journal = {\nat},
         year = 1992,
        month = aug,
       volume = {358},
       number = {6386},
        pages = {477-478},
          doi = {10.1038/358477a0},
       adsurl = {https://ui.adsabs.harvard.edu/abs/1992Natur.358..477P}
}

@ARTICLE{Tavani,
       author = {{Tavani}, M. and {Bulgarelli}, A. and {Piano}, G. and {Sabatini}, S. and {Striani}, E. and {Evangelista}, Y. and {Trois}, A. and {Pooley}, G. and {Trushkin}, S. and {Nizhelskij}, N.~A. and {McCollough}, M. and {Koljonen}, K.~I.~I. and {Pucella}, G. and {Giuliani}, A. and {Chen}, A.~W. and {Costa}, E. and {Vittorini}, V. and {Trifoglio}, M. and {Gianotti}, F. and {Argan}, A. and {Barbiellini}, G. and {Caraveo}, P. and {Cattaneo}, P.~W. and {Cocco}, V. and {Contessi}, T. and {D'Ammando}, F. and {Del Monte}, E. and {de Paris}, G. and {Di Cocco}, G. and {di Persio}, G. and {Donnarumma}, I. and {Feroci}, M. and {Ferrari}, A. and {Fuschino}, F. and {Galli}, M. and {Labanti}, C. and {Lapshov}, I. and {Lazzarotto}, F. and {Lipari}, P. and {Longo}, F. and {Mattaini}, E. and {Marisaldi}, M. and {Mastropietro}, M. and {Mauri}, A. and {Mereghetti}, S. and {Morelli}, E. and {Morselli}, A. and {Pacciani}, L. and {Pellizzoni}, A. and {Perotti}, F. and {Picozza}, P. and {Pilia}, M. and {Prest}, M. and {Rapisarda}, M. and {Rappoldi}, A. and {Rossi}, E. and {Rubini}, A. and {Scalise}, E. and {Soffitta}, P. and {Vallazza}, E. and {Vercellone}, S. and {Zambra}, A. and {Zanello}, D. and {Pittori}, C. and {Verrecchia}, F. and {Giommi}, P. and {Colafrancesco}, S. and {Santolamazza}, P. and {Antonelli}, A. and {Salotti}, L.},
        title = "{Extreme particle acceleration in the microquasar CygnusX-3}",
      journal = {\nat},
         year = 2009,
        month = dec,
       volume = {462},
       number = {7273},
        pages = {620-623},
          doi = {10.1038/nature08578},
archivePrefix = {arXiv},
       eprint = {0910.5344},
 primaryClass = {astro-ph.HE},
       adsurl = {https://ui.adsabs.harvard.edu/abs/2009Natur.462..620T}
}

@ARTICLE{Auger,
       author = {{Pierre Auger Collaboration} and {Abraham}, J. and {Abreu}, P. and {Aglietta}, M. and {Aguirre}, C. and {Allard}, D. and {Allekotte}, I. and {Allen}, J. and {Allison}, P. and {Alvarez}, C. and {Alvarez-Mu{\~n}iz}, J. and {Ambrosio}, M. and {Anchordoqui}, L. and {Andringa}, S. and {Anzalone}, A. and {Aramo}, C. and {Argir{\`o}}, S. and {Arisaka}, K. and {Armengaud}, E. and {Arneodo}, F. and {Arqueros}, F. and {Asch}, T. and {Asorey}, H. and {Assis}, P. and {Atulugama}, B.~S. and {Aublin}, J. and {Ave}, M. and {Avila}, G. and {B{\"a}cker}, T. and {Badagnani}, D. and {Barbosa}, A.~F. and {Barnhill}, D. and {Barroso}, S.~L.~C. and {Bauleo}, P. and {Beatty}, J. and {Beau}, T. and {Becker}, B.~R. and {Becker}, K.~H. and {Bellido}, J.~A. and {BenZvi}, S. and {Berat}, C. and {Bergmann}, T. and {Bernardini}, P. and {Bertou}, X. and {Biermann}, P.~L. and {Billoir}, P. and {Blanch-Bigas}, O. and {Blanco}, F. and {Blasi}, P. and {Bleve}, C. and {Bl{\"u}mer}, H. and {Boh{\'a}cov{\'a}}, M. and {Bonifazi}, C. and {Bonino}, R. and {Boratav}, M. and {Brack}, J. and {Brogueira}, P. and {Brown}, W.~C. and {Buchholz}, P. and {Bueno}, A. and {Busca}, N.~G. and {Caballero-Mora}, K.~S. and {Cai}, B. and {Camin}, D.~V. and {Caruso}, R. and {Carvalho}, W. and {Castellina}, A. and {Catalano}, O. and {Cataldi}, G. and {Caz{\'o}n-Boado}, L. and {Cester}, R. and {Chauvin}, J. and {Chiavassa}, A. and {Chinellato}, J.~A. and {Chou}, A. and {Chye}, J. and {Clark}, P.~D.~J. and {Clay}, R.~W. and {Colombo}, E. and {Concei{\c{c}}{\~a}o}, R. and {Connolly}, B. and {Contreras}, F. and {Coppens}, J. and {Cordier}, A. and {Cotti}, U. and {Coutu}, S. and {Covault}, C.~E. and {Creusot}, A. and {Cronin}, J. and {Dagoret-Campagne}, S. and {Daumiller}, K. and {Dawson}, B.~R. and {de Almeida}, R.~M. and {De Donato}, C. and {de Jong}, S.~J. and {De La Vega}, G. and {de Mello Junior}, W.~J.~M. and {de Mello Neto}, J.~R.~T. and {De Mitri}, I. and {de Souza}, V. and {del Peral}, L. and {Deligny}, O. and {Della Selva}, A. and {Delle Fratte}, C. and {Dembinski}, H. and {Di Giulio}, C. and {Diaz}, J.~C. and {Dobrigkeit}, C. and {D'Olivo}, J.~C. and {Dornic}, D. and {Dorofeev}, A. and {dos Anjos}, J.~C. and {Dova}, M.~T. and {D'Urso}, D. and {DuVernois}, M.~A. and {Engel}, R. and {Epele}, L. and {Erdmann}, M. and {Escobar}, C.~O. and {Etchegoyen}, A. and {Facal San Luis}, P. and {Falcke}, H. and {Farrar}, G. and {Fauth}, A.~C. and {Fazzini}, N. and {Fern{\'a}ndez}, A. and {Ferrer}, F. and {Ferry}, S. and {Fick}, B. and {Filevich}, A. and {Filipcic}, A. and {Fleck}, I. and {Fonte}, R. and {Fracchiolla}, C.~E. and {Fulgione}, W. and {Garc{\'\i}a}, B. and {Garc{\'\i}a G{\'a}mez}, D. and {Garcia-Pinto}, D. and {Garrido}, X. and {Geenen}, H. and {Gelmini}, G. and {Gemmeke}, H. and {Ghia}, P.~L. and {Giller}, M. and {Glass}, H. and {Gold}, M.~S. and {Golup}, G. and {Gomez Albarracin}, F. and {G{\'o}mez Berisso}, M. and {G{\'o}mez Herrero}, R. and {Gon{\c{c}}alves}, P. and {Gon{\c{c}}alves do Amaral}, M. and {Gonzalez}, D. and {Gonzalez}, J.~G. and {Gonz{\'a}lez}, M. and {G{\'o}ra}, D. and {Gorgi}, A. and {Gouffon}, P. and {Grassi}, V. and {Grillo}, A. and {Grunfeld}, C. and {Guardincerri}, Y. and {Guarino}, F. and {Guedes}, G.~P. and {Guti{\'e}rrez}, J. and {Hague}, J.~D. and {Hamilton}, J.~C. and {Hansen}, P. and {Harari}, D. and {Harmsma}, S. and {Harton}, J.~L. and {Haungs}, A. and {Hauschildt}, T. and {Healy}, M.~D. and {Hebbeker}, T. and {Heck}, D. and {Hojvat}, C. and {Holmes}, V.~C. and {Homola}, P. and {H{\"o}randel}, J. and {Horneffer}, A. and {Horvat}, M. and {Hrabovsky}, M. and {Huege}, T. and {Iarlori}, M. and {Insolia}, A. and {Ionita}, F. and {Italiano}, A. and {Kaducak}, M. and {Kampert}, K.~H. and {Keilhauer}, B. and {Kemp}, E. and {Kieckhafer}, R.~M. and {Klages}, H.~O. and {Kleifges}, M. and {Kleinfeller}, J. and {Knapik}, R. and {Knapp}, J. and {Koang}, D. -H. and {Kopmann}, A.},
        title = "{Correlation of the Highest-Energy Cosmic Rays with Nearby Extragalactic Objects}",
      journal = {Science},
         year = 2007,
        month = nov,
       volume = {318},
       number = {5852},
        pages = {938},
          doi = {10.1126/science.1151124},
archivePrefix = {arXiv},
       eprint = {0711.2256},
 primaryClass = {astro-ph},
       adsurl = {https://ui.adsabs.harvard.edu/abs/2007Sci...318..938P}
}

@ARTICLE{Fryer,
       author = {{Fryer}, Chris L. and {Belczynski}, Krzysztof and {Wiktorowicz}, Grzegorz and {Dominik}, Michal and {Kalogera}, Vicky and {Holz}, Daniel E.},
        title = "{Compact Remnant Mass Function: Dependence on the Explosion Mechanism and Metallicity}",
      journal = {\apj},
         year = 2012,
        month = apr,
       volume = {749},
       number = {1},
          eid = {91},
        pages = {91},
          doi = {10.1088/0004-637X/749/1/91},
archivePrefix = {arXiv},
       eprint = {1110.1726},
 primaryClass = {astro-ph.SR},
       adsurl = {https://ui.adsabs.harvard.edu/abs/2012ApJ...749...91F}
}

@ARTICLE{FabianFeedback,
       author = {{Fabian}, A.~C.},
        title = "{Observational Evidence of Active Galactic Nuclei Feedback}",
      journal = {\araa},
         year = 2012,
        month = sep,
       volume = {50},
        pages = {455-489},
          doi = {10.1146/annurev-astro-081811-125521},
archivePrefix = {arXiv},
       eprint = {1204.4114},
 primaryClass = {astro-ph.CO},
       adsurl = {https://ui.adsabs.harvard.edu/abs/2012ARA&A..50..455F}
}

@ARTICLE{IceCube,
       author = {{IceCube Collaboration} and {Aartsen}, M.~G. and {Ackermann}, M. and {Adams}, J. and {Aguilar}, J.~A. and {Ahlers}, M. and {Ahrens}, M. and {Al Samarai}, I. and {Altmann}, D. and {Andeen}, K. and {Anderson}, T. and {Ansseau}, I. and {Anton}, G. and {Arg{\"u}elles}, C. and {Auffenberg}, J. and {Axani}, S. and {Bagherpour}, H. and {Bai}, X. and {Barron}, J.~P. and {Barwick}, S.~W. and {Baum}, V. and {Bay}, R. and {Beatty}, J.~J. and {Becker Tjus}, J. and {Becker}, K. -H. and {BenZvi}, S. and {Berley}, D. and {Bernardini}, E. and {Besson}, D.~Z. and {Binder}, G. and {Bindig}, D. and {Blaufuss}, E. and {Blot}, S. and {Bohm}, C. and {B{\"o}rner}, M. and {Bos}, F. and {B{\"o}ser}, S. and {Botner}, O. and {Bourbeau}, E. and {Bourbeau}, J. and {Bradascio}, F. and {Braun}, J. and {Brenzke}, M. and {Bretz}, H. -P. and {Bron}, S. and {Brostean-Kaiser}, J. and {Burgman}, A. and {Busse}, R.~S. and {Carver}, T. and {Cheung}, E. and {Chirkin}, D. and {Christov}, A. and {Clark}, K. and {Classen}, L. and {Coenders}, S. and {Collin}, G.~H. and {Conrad}, J.~M. and {Coppin}, P. and {Correa}, P. and {Cowen}, D.~F. and {Cross}, R. and {Dave}, P. and {Day}, M. and {de Andr{\'e}}, J.~P.~A.~M. and {De Clercq}, C. and {DeLaunay}, J.~J. and {Dembinski}, H. and {De Ridder}, S. and {Desiati}, P. and {de Vries}, K.~D. and {de Wasseige}, G. and {de With}, M. and {DeYoung}, T. and {D{\'\i}az-V{\'e}lez}, J.~C. and {di Lorenzo}, V. and {Dujmovic}, H. and {Dumm}, J.~P. and {Dunkman}, M. and {Dvorak}, E. and {Eberhardt}, B. and {Ehrhardt}, T. and {Eichmann}, B. and {Eller}, P. and {Evenson}, P.~A. and {Fahey}, S. and {Fazely}, A.~R. and {Felde}, J. and {Filimonov}, K. and {Finley}, C. and {Flis}, S. and {Franckowiak}, A. and {Friedman}, E. and {Fritz}, A. and {Gaisser}, T.~K. and {Gallagher}, J. and {Gerhardt}, L. and {Ghorbani}, K. and {Glauch}, T. and {Gl{\"u}senkamp}, T. and {Goldschmidt}, A. and {Gonzalez}, J.~G. and {Grant}, D. and {Griffith}, Z. and {Haack}, C. and {Hallgren}, A. and {Halzen}, F. and {Hanson}, K. and {Hebecker}, D. and {Heereman}, D. and {Helbing}, K. and {Hellauer}, R. and {Hickford}, S. and {Hignight}, J. and {Hill}, G.~C. and {Hoffman}, K.~D. and {Hoffmann}, R. and {Hoinka}, T. and {Hokanson-Fasig}, B. and {Hoshina}, K. and {Huang}, F. and {Huber}, M. and {Hultqvist}, K. and {H{\"u}nnefeld}, M. and {Hussain}, R. and {In}, S. and {Iovine}, N. and {Ishihara}, A. and {Jacobi}, E. and {Japaridze}, G.~S. and {Jeong}, M. and {Jero}, K. and {Jones}, B.~J.~P. and {Kalaczynski}, P. and {Kang}, W. and {Kappes}, A. and {Kappesser}, D. and {Karg}, T. and {Karle}, A. and {Katz}, U. and {Kauer}, M. and {Keivani}, A. and {Kelley}, J.~L. and {Kheirandish}, A. and {Kim}, J. and {Kim}, M. and {Kintscher}, T. and {Kiryluk}, J. and {Kittler}, T. and {Klein}, S.~R. and {Koirala}, R. and {Kolanoski}, H. and {K{\"o}pke}, L. and {Kopper}, C. and {Kopper}, S. and {Koschinsky}, J.~P. and {Koskinen}, D.~J. and {Kowalski}, M. and {Krings}, K. and {Kroll}, M. and {Kr{\"u}ckl}, G. and {Kunwar}, S. and {Kurahashi}, N. and {Kuwabara}, T. and {Kyriacou}, A. and {Labare}, M. and {Lanfranchi}, J.~L. and {Larson}, M.~J. and {Lauber}, F. and {Leonard}, K. and {Lesiak-Bzdak}, M. and {Leuermann}, M. and {Liu}, Q.~R. and {Lozano Mariscal}, C.~J. and {Lu}, L. and {L{\"u}nemann}, J. and {Luszczak}, W. and {Madsen}, J. and {Maggi}, G. and {Mahn}, K.~B.~M. and {Mancina}, S. and {Maruyama}, R. and {Mase}, K. and {Maunu}, R. and {Meagher}, K. and {Medici}, M. and {Meier}, M. and {Menne}, T. and {Merino}, G. and {Meures}, T. and {Miarecki}, S. and {Micallef}, J. and {Moment{\'e}}, G. and {Montaruli}, T. and {Moore}, R.~W. and {Morse}, R. and {Moulai}, M. and {Nahnhauer}, R. and {Nakarmi}, P. and {Naumann}, U. and {Neer}, G.},
        title = "{Multimessenger observations of a flaring blazar coincident with high-energy neutrino IceCube-170922A}",
      journal = {Science},
         year = 2018,
        month = jul,
       volume = {361},
       number = {6398},
          eid = {eaat1378},
        pages = {eaat1378},
          doi = {10.1126/science.aat1378},
archivePrefix = {arXiv},
       eprint = {1807.08816},
 primaryClass = {astro-ph.HE},
       adsurl = {https://ui.adsabs.harvard.edu/abs/2018Sci...361.1378I}
}

\end{document}